%% file: main.tex
\documentclass[aps, prb, 10pt, twocolumn, groupedaddress, longbibliography, nobalancelastpage, nofootinbib]{revtex4-2}

\input{preamble}

\begin{document}

\title{Efficient MPS methods for extracting spectral information on rings and cylinders}

\author{Maarten Van Damme}
\author{Robijn Vanhove}
\author{Jutho Haegeman}
\author{Frank Verstraete}
\author{Laurens Vanderstraeten}
\affiliation{Department of Physics and Astronomy, University of Ghent, Krijgslaan 281, 9000 Gent, Belgium}

\begin{abstract}
Based on the MPS formalism, we introduce an ansatz for capturing excited states in finite systems with open boundary conditions, providing a very efficient method for computing, e.g., the spectral gap of quantum spin chains. This method can be straightforwardly implemented on top of an existing DMRG or MPS ground-state code. Although this approach is built on open-boundary MPS, we also apply it to systems with periodic boundary conditions. Despite the explicit breaking of translation symmetry by the MPS representation, we show that momentum emerges as a good quantum number, and can be exploited for labeling excitations on top of MPS ground states. We apply our method to the critical Ising chain on a ring and the classical Potts model on a cylinder. Finally, we apply the same idea to compute excitation spectra for 2-D quantum systems on infinite cylinders. Again, despite the explicit breaking of translation symmetry in the periodic direction, we recover momentum as a good quantum number for labeling excitations. We apply this method to the 2-D transverse-field Ising model and the half-filled Hubbard model; for the latter, we obtain accurate results for, e.g., the hole dispersion for cylinder circumferences up to eight sites.
\end{abstract}

\maketitle

\section{Introduction}

\noindent Matrix product states (MPS) \cite{Schollwock2011, Cirac2020}, or the density-matrix renormalization group (DMRG) \cite{White1992, White1993}, provide an efficient formalism for simulating one-dimensional (1-D) and quasi 1-D quantum lattice systems with very high precision. Although MPS are introduced most elegantly on a periodic system \cite{Rommer1997, Verstraete2004, PerezGarcia2006}, MPS and DMRG algorithms are traditionally formulated on finite systems with open boundary conditions, because this setting allows for an efficient calculation of expectation values and the fixing of the gauge degrees of freedom in canonical forms. The disadvantage of open boundaries is that the translation symmetry of the model is explicitly broken, but the formulation of MPS directly in the thermodynamic limit \cite{Ostlund1995, Vidal2007, McCulloch2008, Haegeman2011, Vanderstraeten2019} has made it possible to restore translational symmetries without sacrificing the efficiency or canonical forms of MPS. In this setting, translation symmetry of the MPS ground state has been used as a basis for formulating the MPS version \cite{Haegeman2012} of the Feynman-Bijl ansatz \cite{Feynman1953} or the single-mode approximation \cite{Girvin1985, Arovas1988, Takahashi1994, Sorensen1994}, which describes excited-state wavefunctions with definite momentum quantum number with high precision for generic 1-D lattice models \cite{Bera2017, ZaunerStauber2018b, Vanderstraeten2018}. Localized dynamics on top of such an infinite translation-invariant MPS can be implemented by considering a localized window of different MPS tensors \cite{Milsted2013, Phien2012, Zauner2015}, such that, e.g., spectral functions can be computed for the infinite system \cite{Kjall2011, Gohlke2017}.
\par Yet, systems with periodic boundary conditions are important in, at least, two contexts. First, in the study of 1-D critical models the low-energy spectrum of a finite periodic system provides a very clear fingerprint of the conformal field theory (CFT) that captures the infrared properties of the model \cite{Francesco2012}. Therefore, computing the spectrum on a periodic system is paramount for identifying the effective CFT for a given critical model. Second, the finite-size effects imply that the most efficient way of studying two-dimensional (2-D) quantum systems with MPS consists of imposing periodic boundary conditions in one direction, i.e.\ studying the model on a cylindrical geometry around which the MPS snakes or spirals.
\par In the case of 1-D periodic systems, there have been a number of proposals for alleviating the computational cost of periodic MPS \cite{Porras2006, Pirvu2011, Pirvu2012b, Draxler2017, Zou2018} and DMRG \cite{Pippan2010, Rossini2011} simulations. In addition, a variational ansatz for capturing elementary excitations on top of a periodic MPS was proposed \cite{Pirvu2012a} and refined \cite{Zou2018, Chen2021}, which captures many low-lying energy levels that reflect the CFT finite-size spectrum of critical spin chains \cite{Zou2018}. Here, it is crucial that translation invariance is explicitly conserved in the ground state -- by choosing a uniform MPS ansatz -- such that the momentum is a good quantum number for labeling the excited states.
\par When systems with a cylindrical geometry are studied using a real-space MPS ansatz \cite{Stoudenmire2012}, translation symmetry around the cylinder is explicitly broken by the MPS structure. One can hope, however, that the symmetry is restored in the ground state for sufficiently large bond dimension, and the different modes in the entanglement spectrum can typically be accurately labeled by a transversal momentum \cite{Cincio2013}. Yet, transversal momentum cannot explicitly be used as a good quantum number in the MPS ansatz to target excitations. In the case of fermionic systems, one can switch to the momentum basis in the periodic direction such that translation symmetry can be preserved explicitly \cite{Motruk2016, Ehlers2017}, but in the case of spin systems the transformation to the transversal momentum basis cannot be implemented as a canonical transformation and does not exhibit a tensor-product structure.

\par In this paper, we show that we can recover momentum as a reliable quantum number to label excitations even when the associated translation symmetry is explicitly broken by the MPS structure. In particular, we will use open-boundary MPS and consider its tangent space as a variational ansatz for excitations, analogously to the excitation ansatz that has proven successful for infinite systems. First, we set the stage by applying this ansatz to an actual finite 1-D spin system, namely to study the magnon excitations in the spin-1 Heisenberg model, and compare the results to traditional DMRG-based approaches for targeting excited states. Next, we apply the same formalism to 1-D systems with periodic boundary conditions, i.e.\ living on a ring, where we find that momentum is recovered with good accuracy. This is useful in particular for studying critical systems, and we obtain very accurate CFT spectra for Ising and Potts models. Finally, we translate the same idea to capturing excitations in infinite cylinders. Again, despite the explicit breaking of translation symmetry around the cylinder, the transversal momentum can be recovered accurately and we discuss a strategy to target specific momentum sectors. We illustrate this approach for the 2-D transverse-field Ising model and the half-filled 2-D Hubbard model. All algorithms in this paper can be found in \emph{MPSKit.jl} \cite{mpskit}, an open-source software package written in the scientific programming language Julia.

\section{Quasiparticles on a finite system}

There are a variety of techniques to variationally target excited states in traditional DMRG or MPS simulations. The easiest is when the excitation has a non-trivial quantum number, such that it can directly be obtained as the solution of a ground state problem in that symmetry sector \cite{White1993}. If multiple states in the same sector are required, they can be simultaneously captured by an extended DMRG procedure \cite{White1993, Schollwock2005}, but the bond dimension grows rapidly with the number of targeted states. In the MPS representation, one would instead sequentially find higher excited states by imposing orthogonality on previously found states \cite{McCulloch2007}. Finally, in the case of critical systems, it was realized recently that just changing one tensor in the middle of the chain is often sufficient to capture the low-lying excitations \cite{Chepiga2017}.
\par In this section, we introduce a variational ansatz for directly capturing the excited states of finite quantum chains and show that we can outperform these standard approaches, both in efficiency and accuracy.

\subsection{Method}

\noindent Let us consider a finite one-dimensional spin chain of length $N$ with open boundary conditions, described by a generic model Hamiltonian $H$, for which the ground state is described by an MPS
\begin{equation} \label{eq:mps}
\ket{\Psi(A_1\dots A_N)} = \diagram{p1}{1}.
\end{equation}
This MPS representation has a residual gauge freedom. In particular, we can always choose the tensors left and right of a given site to be respectively left and right isometries, leading to the representation
\begin{equation}
\ket{\Psi(A_1\dots A_N)} = \diagram{p1}{2}.
\end{equation}
with $A^l_i$ and $A^r_i$ the left and right isometric MPS tensors, and $A^c_i$ the center-site tensor; this center site can be chosen anywhere in the chain by performing left- and right-isometric decompositions on the MPS tensors. The standard sweeping algorithm \cite{Schollwock2011} can be used to find an optimal MPS approximation for the ground state.
\par Once we have found a good MPS approximation of the ground state, we can build excitations on top. Inspired by the success of the quasiparticle ansatz in the thermodynamic limit, we propose a very similar ansatz
\begin{equation}
\ket{\Phi(B_1,\dots,B_N)} = \sum_i \diagram{p1}{3}.
\end{equation}
In this ansatz, we take a sum of $N$ terms, where in each term we bring the MPS into canonical centered around site $i$ and we modify this one center-site tensor. Note that one can represent this state as a finite MPS with two times the bond dimension of the ground state.
\par The quasiparticle ansatz exhibits a gauge freedom, since modifying all tensors $B_i$ as $B'_i = B_i + Y_i A_i^r - A_i^l Y_{i+1}$ with a set of matrices $\{Y_i\}$ simultaneously would leave the state invariant. We can follow the ideas from infinite MPS \cite{Vanderstraeten2019}, and fix this gauge freedom by imposing that $B_i$ lives in the null-space of $(A^l_i)^\dagger$. The tensor $B_i$ is then parametrized as 
\begin{equation} \label{eq:gauge_fix}
\diagram{p1}{4} = \diagram{p1}{5}
\end{equation}
where $X_i$ contains the actual degrees of freedom and $V_i$ spans the null space of $(A^l_i)^\dagger$, i.e. it contains the orthonormal columns to complement $A^l_i$ to a full unitary matrix. Diagrammatically, the $V_i$ tensor satisfies the conditions\footnote{Close to the edge of the system, the $A_l$ and $A_r$ tensors are full-rank unitaries instead of isometries. Therefore, on the left edge of the chain the null spaces $V_i$'s are empty, and no $X_i$'s are introduced. On the right edge, the $X_i$'s are non-zero.}
\begin{equation}
\diagram{p1}{6} = 0, \qquad \diagram{p1}{7}=\diagram{p1}{8}.
\end{equation}
As a consequence of this gauge fixing, the quasiparticle state is automatically orthogonal to the ground-state MPS. Our ansatz now becomes
\begin{multline}
\ket{\Phi(X_1\dots X_N} = \sum_i \ket{\Phi_i(X_i)},  \\
\ket{\Phi_i(X_i)} = \diagram{p1}{9}.
\end{multline}
The overlap of two distinct quasiparticle states simplifies to the simple euclidean inner product\footnote{Here we denote $\vec{X}_i$ for the vectorized version of the tensor $X_i$, and $\vec{X}$ for the concatenation of the vectors $\vec{X}_i$ into a large vector.} of the tensors $X_i$
\begin{align}
\braket{ \Phi(X_1\dots X_N) | \Phi(X_1'\dots X_N') } &= \sum_i (\vec{X_i})\dag \vec{X_i}' \nonumber \\
&= (\vec{X}')\dag \vec{X}
\end{align}
This is important, as minimizing the energy within the manifold of quasiparticle excitations now becomes a simple eigenvalue problem
\begin{equation}
\sum_{j} (H_{\mathrm{eff}})_{ij} \vec{X}_j = \omega \vec{X}_i,
\end{equation}
with the effective hamiltonian matrix
\begin{equation}
 (\vec{X}_i)\dag (H_{\mathrm{eff}})_{ij} \vec{X}_j = \bra{\Phi_i(X_i)} H \ket{\Phi_j(X_j)}.
\end{equation}
The action of $H_{\mathrm{eff}}$ on a set of tensors $X_i$ can be computed efficiently, as we show explicitly in the Appendix \ref{sec:appendix}, and the eigenvalue problem can be solved by an iterative Krylov method, typically the Lanczos method.

\subsection{Magnon in the spin-1 chain}

\begin{figure}
\includegraphics[width=0.99\columnwidth]{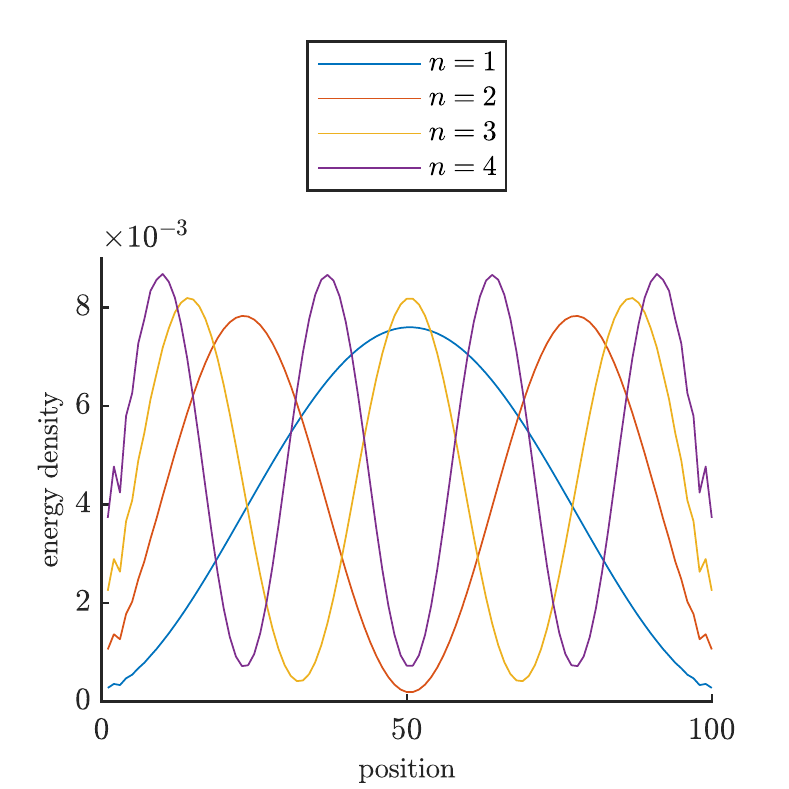}
\caption{The energy density of the four lowest-lying spin-1 magnon excitations on top of the ground state of the spin-1 Heisenberg chain with 100 sites. The ground state has bond dimension $D=64$.}
\label{fig:heisenberg}
\end{figure}

\noindent In infinite systems, the excitation ansatz leads to a picture of gapped excitations as dressed quasiparticles against a correlated background state \cite{Vanderstraeten2015}; in that setting the ansatz describes a traveling wave or Bloch wave with a definite momentum. On a finite system with open boundary conditions, however, we would expect to find a standing-wave configuration of this very same dressed particle.
\par As an illustration of this scenario, we simulate the excitation spectrum of the spin-1 Heisenberg chain on a finite system with open boundary conditions. In order to focus on the bulk excitations, we place spin-1/2s at the ends, thus eliminating the gapless edge modes \cite{White1993b}. In Fig.~\ref{fig:heisenberg} we plot the energy density of the first five excited states, showing indeed the different standing-wave patterns of the magnon behaving as a particle-in-a-box \cite{Sorensen1993}.
\par We can compare this method for extracting excited states with the conventional MPS methods. As explained above, the standard method requires an entire sweeping optimization for every excited state, and one needs higher bond dimensions to faithfully capture the excitation. In that respect, the quasiparticle ansatz is numerically much cheaper as it only requires to solve a single eigenvalue problem for a given number of excited states. Moreover, the excitation ansatz reaches the same level of accuracy for the energy, as we show explicitly in Table \ref{table:heisex}. One expects the excitation ansatz to fail for higher excitations, as soon as they start to involve multiple particles. It then makes sense to go to a hybrid setup where we switch to DMRG after the first few lowest eigenvectors. This will not be necessary however when doing simulations of a system close to criticality and for which the correlation length of the MPS is larger than the system size (i.e. the finite size scaling regime): in that case, the local tensors in the quasiparticle ansatz have a global effect, and seem to be able to represent multiparticle excitations \cite{Chepiga2017}.

\begin{table}

\begin{center}
 \begin{tabular}{|c c c c|}
 \hline
 Energy QP & Energy DMRG & Variance QP & Variance DMRG \\
 \hline\hline
0.4165739 & 0.4165864 & 3.1e-6 & 7.05e-5 \\
\hline
0.4165739 & 0.4165865 & 3.1e-6 & 7.09e-5 \\
\hline
0.4165739 & 0.4165864 & 3.1e-6 & 7.05e-5 \\
\hline
0.4344129 & 0.4344322 & 3.2e-6 & 0.00011 \\
\hline
0.4344129 & 0.4344321 & 3.2e-6 & 0.000109 \\
\hline
0.4344129 & 0.4344321 & 3.2e-6 & 0.000109 \\
\hline
0.462799 & 0.4628202 & 3.3e-6 & 0.0001207 \\
\hline
0.462799 & 0.4628204 & 3.3e-6 & 0.0001209 \\
\hline
0.462799 & 0.4628205 & 3.3e-6 & 0.0001209 \\
\hline
0.5001103 & 0.5001329 & 3.3e-6 & 0.000129 \\
\hline
\end{tabular}
\end{center}
\caption{Comparison of the energies and energy variances obtained using the quasiparticle (QP) ansatz and using DMRG, for the lowest-lying excitations in the spin-1 chain with $N$ sites (also see Fig.~\ref{fig:heisenberg}). For the DMRG excitations, the maximum number of sweeps was set at 100.}
\label{table:heisex}
\end{table}

\section{Critical systems on a ring}

\noindent As highlighted above, MPS techniques are significantly less efficient for systems with periodic boundary conditions, in part due do the inferior scaling in bond dimension when using an MPS with periodic boundary conditions. Still, both ground states and excited states can be targeted with high precision using periodic MPS. 
\par An alternative approach is to re-use the above technique for open boundary systems, but with a periodic hamiltonian or transfer matrix. Although it would require a quadratically larger bond dimension to represent a periodic MPS as an open-boundary MPS, we will show that we obtain quantitatively good results at reasonable bond dimensions. In particular, we can compute the expectation value of the translation operator, which for the MPS ground state [Eq.~\ref{eq:mps}] is given by 
\begin{equation}
\bra{\Psi(A_1\dots)} T \ket{\Psi(A_1 \dots)} = \diagram{ring}{1},
\end{equation}
and show that the momentum of the ground states, as well as the excited states\footnote{Eigenvalues occur with different degeneracies and the corresponding eigenvectors will be momentum superpositions.  We then have to diagonalize the translation operator within this degenerate-energy subspace to extract the momentum labels. It is also possible to work the other way around, and impose a certain momentum. That is precisely what we do in the next section, for quantum systems on a cylinder.}, are correct up to very high precision.
\par In Ref.~\onlinecite{Zou2018} it was shown that the excitation ansatz for periodic MPS is able to reproduce a surprising number of energy levels in the finite-size CFT spectrum of critical spin chains -- even the multi-particle excitations are well captured by this ansatz. In order to make this possible, the bond dimension of the ground-state MPS needs to be large enough such that we are in the finite-size scaling regime \cite{Pirvu2012b}: through the virtual level of the MPS one can change the state over the whole system by modifying a single tensor locally. This same effect was observed in Ref.~\onlinecite{Chepiga2017} for systems with open boundary conditions. Motivated by this effect, we now apply our open-boundary MPS excitation ansatz to critical models on a ring.

\subsection{Quantum Ising chain}

\begin{figure}
\includegraphics[width=0.99\columnwidth]{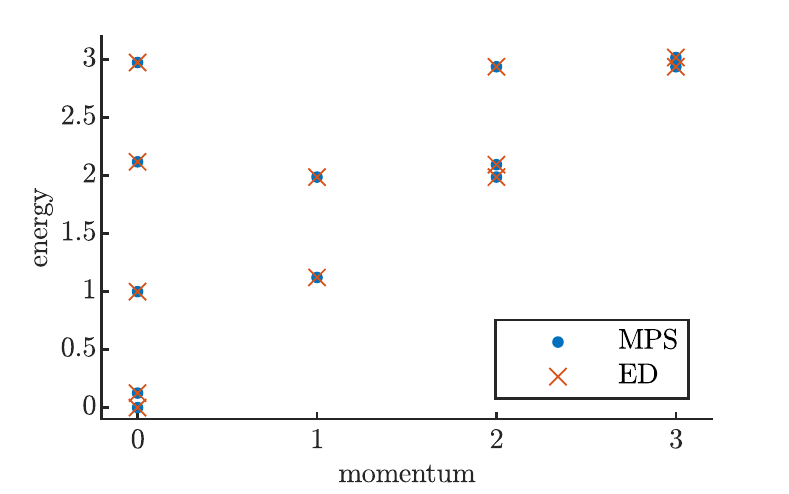}

\caption{The lowest-lying energy levels for the critical quantum Ising chain on a circle with circumferences $N=20$. The blue dots are results from the MPS excitation ansatz with $D=50$ and the red crosses are results from exact diagonalization. The energies were shifted and rescaled such that the ground state is at $e_0=0$ and the gap $\Delta_\epsilon=1$. We see that the MPS with open boundary conditions restores translation invariance, and that degeneracies agree with exact results.}
\label{fig:qcft_ising}
\end{figure}

\begin{table}

\begin{center}
 \begin{tabular}{|c c c c c|}
 \hline
 20 & 30 & 40 & 50 & CFT \\
 \hline\hline
0.0 & 0.0 & 0.0 & 0.0 & 0.0 \\
\hline
0.125 & 0.125 & 0.125 & 0.125 & 0.125 \\
\hline
0.998458 & 0.999314 & 0.999615 & 0.999757 & 1.0 \\
\hline
1.12038 & 1.12295 & 1.12384 & 1.12426 & 1.125 \\
\hline
1.12038 & 1.12295 & 1.12384 & 1.12427 & 1.125 \\
\hline
1.98462 & 1.99316 & 1.99615 & 1.99754 & 2.0 \\
\hline
1.98462 & 1.99316 & 1.99615 & 1.99758 & 2.0 \\
\hline
1.98462 & 1.99316 & 1.99616 & 1.9976 & 2.0 \\
\hline
1.98462 & 1.99316 & 1.99617 & 1.99763 & 2.0 \\
\hline
2.09125 & 2.10996 & 2.11653 & 2.11958 & 2.125 \\
\hline
2.09125 & 2.10996 & 2.11653 & 2.11959 & 2.125 \\
\hline
2.11576 & 2.12089 & 2.1227 & 2.1236 & 2.125 \\
\hline
2.93422 & 2.97063 & 2.98346 & 2.98944 & 3.0 \\
\hline
2.93422 & 2.97063 & 2.98346 & 2.98945 & 3.0 \\
\hline
2.93422 & 2.97063 & 2.98348 & 2.98949 & 3.0 \\
\hline
2.93422 & 2.97064 & 2.98352 & 2.98961 & 3.0 \\
\hline\hline
\end{tabular}
\end{center}
\caption{Comparison of the energies obtained using the quasiparticle (QP) ansatz, for the lowest-lying excitations in the quantum ising chain with $N$ sites. We see convergence towards the predicted CFT results.}
\label{table:isex}
\end{table}

\noindent Let us first look at the simplest critical 1-D model, the critical Ising chain.  As shown in Fig.~\ref{fig:qcft_ising}, our results agree well with exact diagonalization. Despite the open boundary conditions, the MPS restores translation invariance and we retrieve momentum eigenstates. Because our method does not scale exponentially in system size, we can push the simulation to much larger system sizes. Indeed, Tab.~\ref{table:isex} shows that we can approximate the CFT prediction to higher precision by reaching larger systems. This is important for systems with a large local dimension, where one often cannot reach large enough sizes to perform finite-size scaling.

\subsection{Classical 2-D Potts model}

\begin{figure}
\includegraphics[width=0.99\columnwidth]{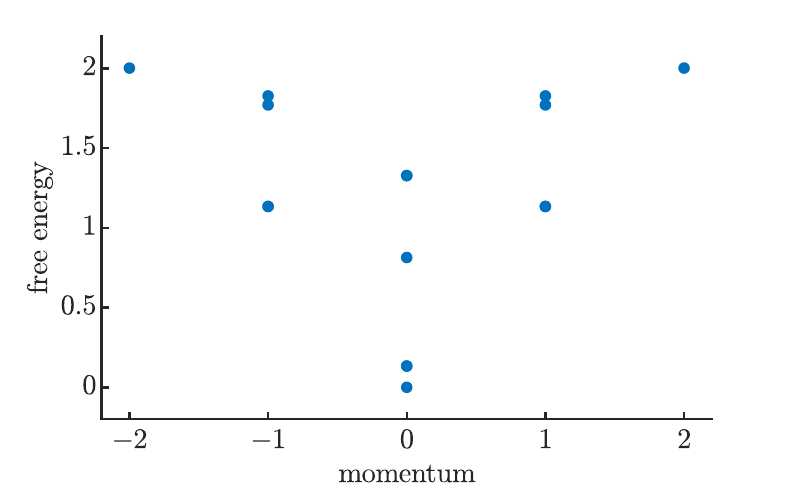}
\caption{The rescaled spectrum of the transfer matrix of the 3-state Potts model with system size $N=28$, as obtained with the excitation ansatz with bond dimension $D=150$. We take the negative logarithm of the transfer-matrix eigenvalues, $f=-\log\lambda$, such that $f$ corresponds to a free energy. We have rescaled the values such that fixed-point free energy is set at $f=0$ and the first excited state at $f=2/15$.}
\label{fig:cft_potts}
\end{figure}

\noindent We can also deal with statistical-mechanical problems on an infinite cylinder, by finding the leading eigenvectors of the transfer matrix \cite{Haegeman2017} in the periodic direction. The largest few eigenvalues contain information about the free energy and the dominant correlation functions in the system. It is possible to find the dominant eigenvector by modifying any of the well known ground-state algorithms to look for the largest-magnitude eigenvalue, and the above algorithm for finding the excited states of a quantum Hamiltonian can be straightforwardly adapted to the transfer-matrix setting.
\par Here we show the finite-size spectrum of the critical 3-state Potts model. In Fig.~\ref{fig:cft_potts} we show the spectrum of the transfer matrix with circumference $N=28$ -- a system size that is not feasible with exact diagonalization -- at bond dimension $D=150$. The rescaling was done such that the ground state sits at free energy $f=0$ and the first excited state at $f=2/15$ (the smallest non-trivial scaling dimension for the Potts CFT); all other eigenvalues are approaching the CFT prediction. In Tab.~\ref{table:cftpotts} we compare the obtained energies with both the exact results on smaller systems and the CFT prediction for the infinite-size limit. An alternative rescaling could be performed, where the free energy at momentum $2$ is rescaled such that it lies at free energy $f=2$ (the first excited state of the identity tower), exploiting the momentum information explicitly. Such a rescaling is independent of the CFT \cite{Francesco2012}, but looking at Tab.~\ref{table:cftpotts} at $N=28$, it makes almost no difference compared to the rescaling that we have used.

\begin{table}
\begin{center}
 \begin{tabular}{|c c c c c|}
 \hline
 10 & 12 & 14 & 28 & CFT \\
 \hline\hline
 0.0 & 0.0 & 0.0 & 0.0 & 0.0 \\
 \hline
 0.13333 & 0.13333 & 0.13333 & 0.13333 & 0.13333 \\
 \hline
 0.13333 & 0.13333 & 0.13333 & 0.13333 & 0.13333 \\
 \hline
 0.83376 & 0.82863 & 0.82499 & 0.8139 & 0.8 \\
 \hline
 1.148 & 1.1428 & 1.1398 & 1.134 & 1.1333 \\
 \hline
 1.148 & 1.1428 & 1.1398 & 1.134 & 1.1333 \\
 \hline
 1.148 & 1.1428 & 1.1398 & 1.134 & 1.1333 \\
 \hline
 1.148 & 1.1428 & 1.1398 & 1.134 & 1.1333 \\
 \hline
 1.3213 & 1.3225 & 1.3235 & 1.3277 & 1.3333 \\
 \hline
 1.3213 & 1.3225 & 1.3235 & 1.3277 & 1.3333 \\
 \hline
 1.7462 & 1.75 & 1.7537 & 1.7704 & 1.8 \\
 \hline
 1.7462 & 1.75 & 1.7537 & 1.7704 & 1.8 \\
 \hline
 1.8995 & 1.8763 & 1.8618 & 1.8272 & 1.8 \\
 \hline
 1.8995 & 1.8763 & 1.8618 & 1.8272 & 1.8 \\
 \hline
 2.0532 & 2.0336 & 2.0224 & 2.0016 & 2.0 \\
 \hline
 2.0532 & 2.0336 & 2.0224 & 2.0017 & 2.0 \\
 \hline
\end{tabular}
\end{center}
 \caption{Comparison of the rescaled free energies of the 3-state Potts model at different system sizes with the CFT prediction. Results for system sizes $N=(10,12,14)$ were obtained with exact diagonalization, while results for system size $N=28$ where obtained with our MPS-based excitation ansatz with bond dimension $D=150$.}
 \label{table:cftpotts}
\end{table}

\section{Systems on a cylinder}

\noindent We can generalize this approach to the setting of MPS approximations for cylindrical systems, a setup that is very often used for simulating 2-D systems. Here, the MPS is wrapped around the infinite cylinder in a snake-like fashion, so the translation symmetry in the transversal direction is explicitly broken by the MPS representation. However, just like for the one-dimensional rings above, we can hope that this translation symmetry is restored for large enough bond dimensions and that we can exploit this to create quasiparticle excitations with fixed transversal momentum. In the case of infinite cylinders, translation symmetry along the cylinder can be imposed straightforwardly, so in this way we can have access to the momentum quantum numbers in both directions.

\subsection{Method}

\noindent The ground state of an infinite cylinder with a circumference of $N$ sites can be represented by an infinite MPS with an $N$-site unit cell\footnote{The analysis is easily generalized to larger unit cells.}
\begin{equation}
\ket{\Psi(\{A_i\})} = \diagram{cylinder}{1}.
\end{equation}
This MPS can be found using variational ground-state searches \cite{ZaunerStauber2018, Vanderstraeten2019} or the infinite DMRG algorithm \cite{McCulloch2008}. We normalize the state such that the leading eigenvalue of the $N$-site transfer matrix is one,
\begin{equation}
\lambda_{\mathrm{max}}\left( \diagram{cylinder}{2} \right) = 1.
\end{equation}
There are two translation operators acting on this state. The first one ($T_x$) corresponds to translation along the cylinder, and shifts the full unit cell
\begin{equation}
T_x \ket{\Psi(\{A_i\})} = \diagram{cylinder}{3}.
\end{equation}
The second one ($T_y$) corresponds to translation around the cylinder and is a transformation within the unit cell,
\begin{equation}
T_y \ket{\Psi(\{A_i\})} = \diagram{cylinder}{4}.
\end{equation}
The MPS is invariant under $T_x$ by construction, but translation invariance under $T_y$ is less trivial. The expectation value of this operator
\begin{equation}
\bra{\Psi(\{A_i\})} T_y \ket{\Psi(\{A_i\})},
\end{equation}
is a quantity that scales exponentially with the number of unit cells, so the characteristic quantity is the leading eigenvalue of the mixed $N$-site transfer matrix,
\begin{equation} \label{eq:mu}
\mu = \lambda_{\mathrm{max}} \left( \diagram{cylinder}{5} \right),
\end{equation}
so that, formally, $\bra{\Psi(\{A_i\})} T_y \ket{\Psi(\{A_i\})} \sim \mu^{N_x}$ with $N_x$ the diverging number of unit cells. For future reference, we can associate left and right fixed points to this mixed transfer matrix as
\begin{align}
\diagram{cylinder}{6} = \mu \diagram{cylinder}{7} \\
\diagram{cylinder}{8} = \mu \diagram{cylinder}{9}.
\end{align}
Now the MPS has well-defined transversal momentum if the eigenvalue $\mu$ lies on the unit circle with the angle a multiple of $2\pi/N$.

\par We can make an excitation on top of this MPS with the ansatz
\begin{multline}
\ket{\Phi_{p_x}(B)} = \sum_{n} \e^{ip_xn} T_x^n \\ \diagram{cylinder}{10}  
\end{multline}
with
\begin{multline}
\diagram{cylinder}{11} = \diagram{cylinder}{12} + \dots \\ \dots + \diagram{cylinder}{13}.
\end{multline}
This is the multi-site version \cite{ZaunerStauber2018b} of the excitation ansatz \cite{Haegeman2012} for infinite MPS. Similar to Eq.~\ref{eq:gauge_fix}, we can fix the redundant gauge degrees of freedom in the $B$ tensors such that the overlap between two of these excited states reduces to the simple euclidean inner product with a $\delta$-function normalization for the momentum,
\begin{equation}
\braket{\Phi_{p_x'}(B') | \Phi_{p_x}(B)} = 2\pi\delta(p_x-p_x') \sum_{i=1}^N (\vec{B}_i')^\dagger \vec{B}_i .
\end{equation}
Consequently, an ordinary eigenvalue problem
\begin{equation}
H_{p_x,\mathrm{eff}} \vec{B} = \omega \vec{B},
\end{equation}
with
\begin{multline}
\bra{\Phi_{p_x'}(B')} H \ket{\Phi_{p_x}(B)} \\ = 2\pi\delta(p_x-p_x') (\vec{B})\dag H_{p_x,\mathrm{eff}} \vec{B} ,
\end{multline}
finds the optimal $B$ tensors for representing the lowest-energy excitation in the system. 
\par This ansatz clearly is an eigenstate of the $T_x$ operator,
\begin{equation}
T_x\ket{\Phi_{p_x}(B)} = \e^{ip_x} \ket{\Phi_{p_x}(B)},
\end{equation}
but the momentum around the cylinder is not straightforward. Indeed, we can compute the normalized expectation value for the $T_y$ operator by dividing by the ground-state expectation value 
\begin{align}
&  \frac{1}{2\pi\delta(p_x-p_x')} \frac{\bra{\Phi_{p_x'}(B)} T_y \ket{\Phi_{p_x}(B)}}{\bra{\Psi(A)} T_y \ket{\Psi(A)} } \nonumber \\
& \hspace{1cm} = \frac{1}{\mu} \diagram{cylinder}{14} \nonumber \\
& \hspace{1.8cm} + \frac{\e^{-ip_x}}{\mu^2} \diagram{cylinder}{15} \nonumber \\
& \hspace{1.8cm} + \dots \;,
\end{align}
The infinite sums in this expression converge so that this expression yields a finite number; we can hence compute the transversal momentum of the excited states that we find variationally. We can, however, directly target excitations that are approximate eigenvectors of the $T_y$ operators\footnote{In a recent work \cite{Wang2019}, a similar MPS approach was proposed for targeting a given value of the momentum in a spin chain.} Indeed, if we define the effective $T_y$ operator as  
\begin{equation}
 \frac{\bra{\Phi_{p_x}(B')} T_y \ket{\Phi_{p_x}(B)}}{\bra{\Psi(A)} T_y \ket{\Psi(A)} } = 2\pi\delta(p_x-p_x') (\vec{B}')\dag T_{\mathrm{eff}} \vec{B} ,
\end{equation}
we solve the eigenvalue problem 
\begin{equation} \label{eq:eig_momentum}
\left( H_{p_x,\mathrm{eff}} - \alpha \e^{-ip_y} T_{\mathrm{eff}} \right) B_{p_x,p_y} = \lambda B_{p_x,p_y}
\end{equation}
for the eigenvalue $\lambda = \omega - \alpha$ with the most negative real part.\footnote{It would be more elegant to consider the operator
\begin{equation}
H_{p_x,\mathrm{eff}} - \alpha \e^{-ip_y} T_{\mathrm{eff}} - \alpha \e^{ip_y} T_{\mathrm{eff}} \dag,
\end{equation}
such that the eigenvalue problem remains hermitian. In practice, however, we find that adding the hermitian conjugate is not needed for the stability of the eigenvalue problem and only increases the computational cost.} When $\alpha$ is sufficiently large\footnote{The numerical value of $\alpha$ depends on the problem. In practice, one computes the transversal momentum of the obtained state to check whether $\alpha$ was indeed chosen large enough; if not, one should restart the optimization with a larger value of $\alpha$.} and translation symmetry is sufficiently well captured by the MPS, this eigenvalue should indeed be real and correspond to an eigenstate with transversal momentum $p_y$ and energy eigenvalue $\omega$.

\subsection{Benchmarks}

\begin{figure}
\includegraphics[width=0.99\columnwidth]{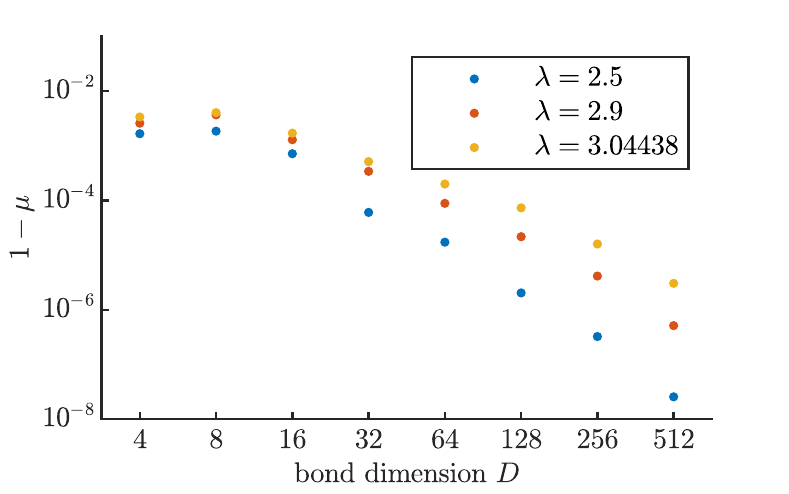}
\caption{The momentum per rung of the MPS ground state [Eq.~\ref{eq:mu}] for the square-lattice Ising model on an $N=12$ cylinder as a function of bond dimension for three values of the field, $\lambda=2.5$ (blue), $\lambda=2.9$ (red) and $\lambda=3.04438$ (orange).}
\label{fig:mom_ising}
\end{figure}

\begin{figure}
\subfigure{\includegraphics[width=0.99\columnwidth]{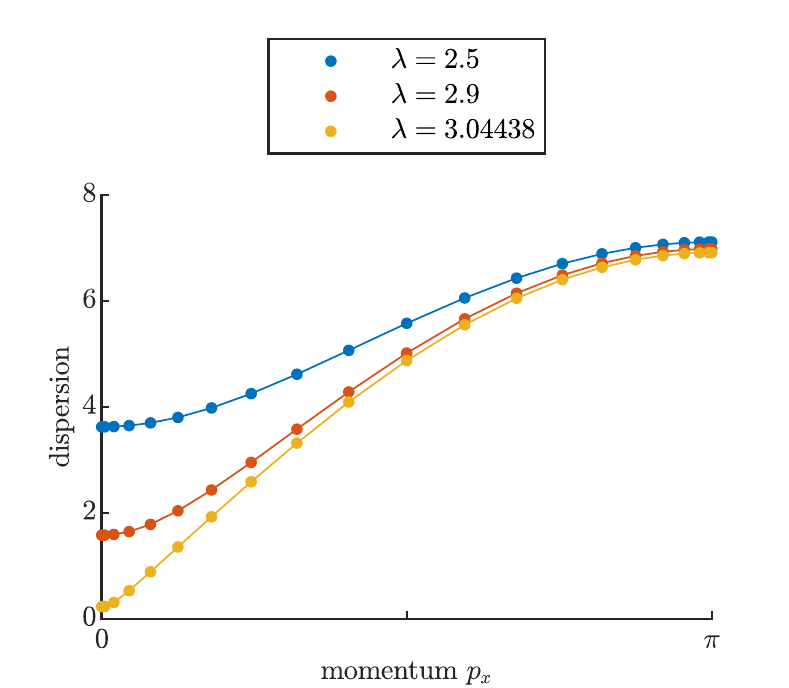}} \\
\subfigure{\includegraphics[width=0.99\columnwidth]{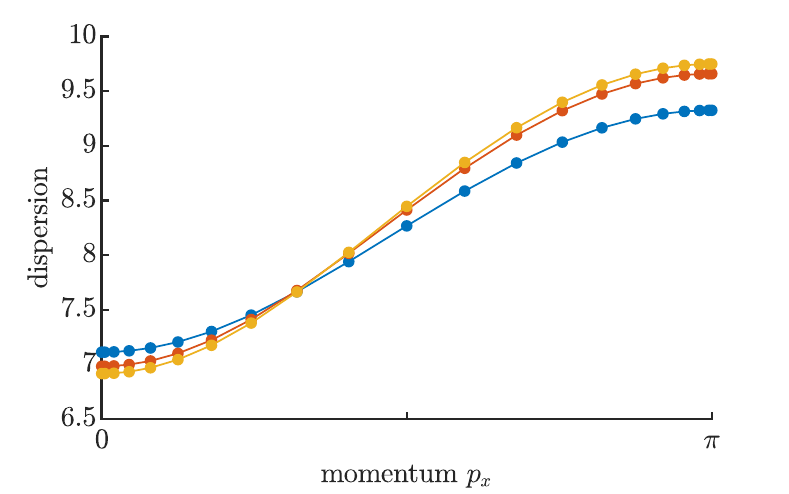}}
\caption{The dispersion relation of the square-lattice Ising model on a cylinder of circumference $N=12$ for three values of the field, $\lambda=2.5$ (blue), $\lambda=2.9$ (red) and $\lambda=3.04438$ (orange). The top panel show the BZ cut at transversal momentum $p_y=0$, nicely capturing the spectrum becoming gapless at the critical point. The bottom panel is for $p_y=\pi$, where the dispersion changes very little as the field is tuned. Here, all simulations were done with an MPS bond dimension $D=256$.}
\label{fig:disp_ising}
\end{figure}

\begin{figure}
\includegraphics[width=0.99\columnwidth]{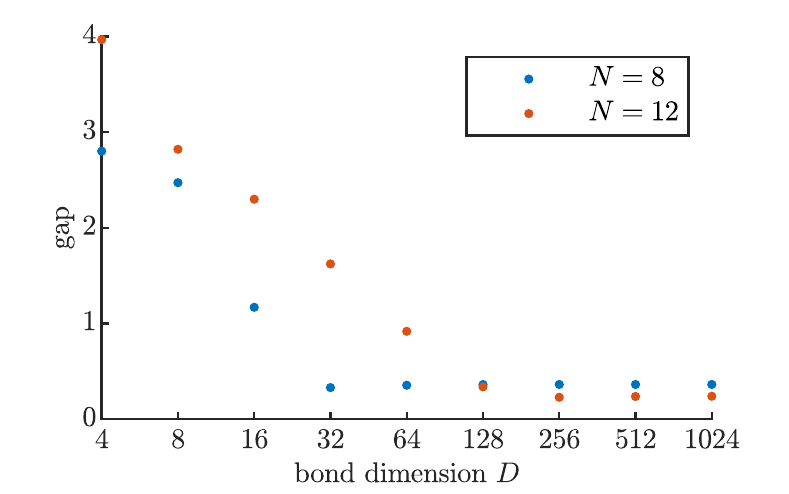}
\caption{The gap at $\lambda=3.04438$ for cylinders $N=8$ (blue) and $N=12$ (red). We can converge the value of the gap with increasing bond dimension in both cases, although the larger cylinder needs higher bond dimension. The non-zero gap is due to the finite circumference of the cylinder, and clearly goes down as $N$ increases.}
\label{fig:gap_ising}
\end{figure}

\noindent We first illustrate this approach on the two-dimensional transverse-field Ising model, defined by the Hamiltonian
\begin{equation}
H_{\text{Ising}} = \sum_{\braket{ij}} S^z_i S^z_j + \lambda \sum_i S^x_i.
\end{equation}
We choose to run simulations at two values in the symmetry-broken phase, $\lambda=2.5$ and $\lambda=2.9$, and at the quantum critical point $\lambda=3.04438$ (taken from Ref.~ \onlinecite{Blote2002}).
\par Let us first consider the momentum of the MPS ground state. We have optimized MPS ground states on infinite cylinders with circumference $N=12$ at different bond dimensions; Fig.~\ref{fig:mom_ising} shows that the expectation value of $T_y$ per rung [Eq.~\ref{eq:mu}] nicely converges to one. The convergence is slower as one approaches the critical point, which points to the fact that a larger bond dimension is needed near criticality.

\par Next we study the excitations, where we can take different cuts of the two-dimensional Brillouin zone by imposing the transversal momentum as in Eq.~\eqref{eq:eig_momentum}, see Fig.~\ref{fig:disp_ising}. The $p_y=0$ cut shows the dispersion becoming gapless at the critical point, whereas the $p_y=\pi$ cut is gapped. We push the bond dimension to get a good estimate for the gapless point, showing that the finite circumference induces a non-zero gap even for the critical value of the field; only in the infinite 2-D plane the model becomes truly gapless. As we show in Fig.~\ref{fig:gap_ising}, the finite-size gap decreases for increasing cylinder width.

\begin{figure}
\subfigure{\includegraphics[width=0.99\columnwidth]{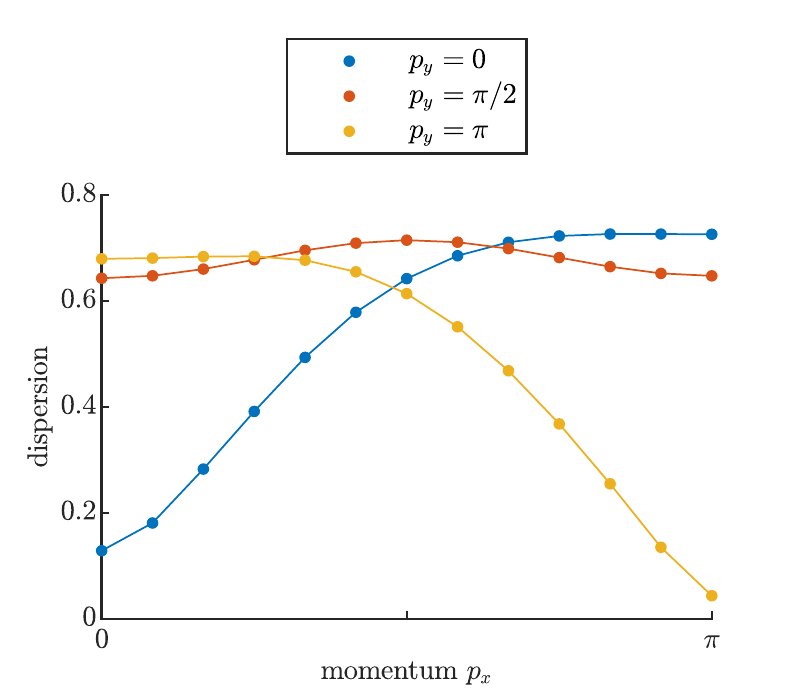}} \\
\subfigure{\includegraphics[width=0.99\columnwidth]{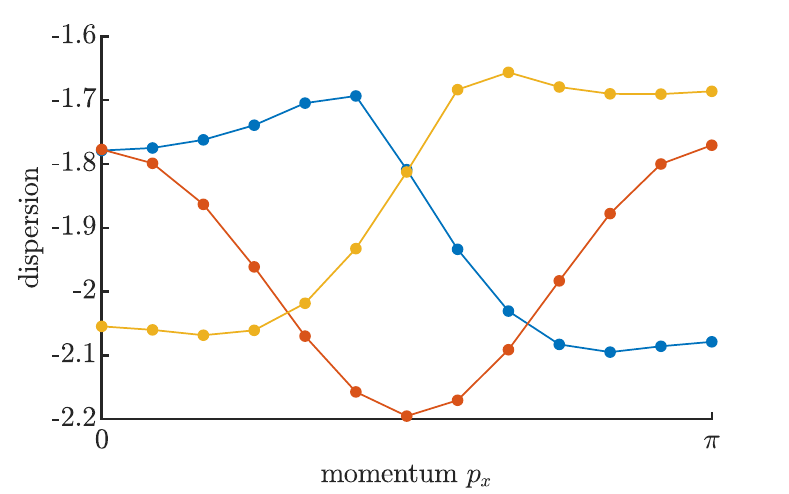}}
\caption{The dispersion relation of the Hubbard model with $t=1$ and $U=12$ on a cylinder of circumference $N=4$, in three cuts of the Brillouin zone $p_y=0$ (blue), $p_y=\pi/2$ (red) and $p_y=\pi$ (orange). The top panel shows the dispersion of the magnon excitation (charge-neutral spin-1 excitation) in three momentum cuts, and the bottom panel shows the hole dispersion (charged spin-1/2 excitation). Here, the total bond dimension is around $D=3500$, where the largest $\SU(2)\otimes\U(1)$ symmetry block is $D_{\mathrm{max}}=154$.}
\label{fig:disp_hubbard}
\end{figure}

\begin{figure}
\includegraphics[width=0.99\columnwidth]{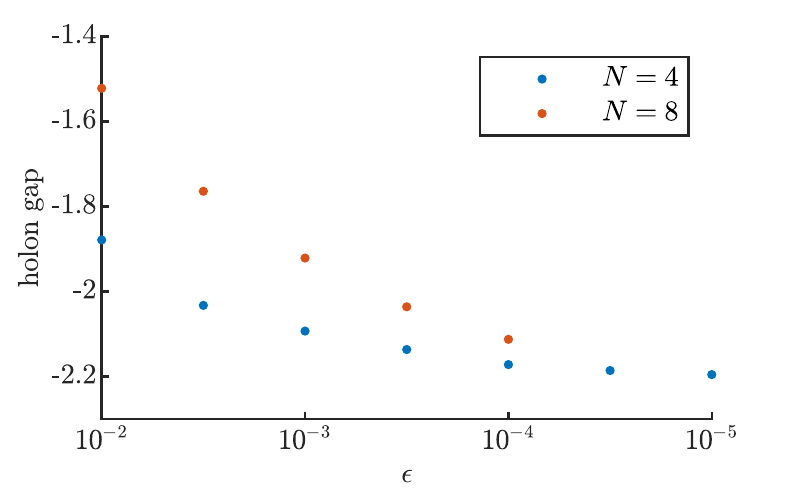}
\caption{The minimum of the dispersion relation at the S point (momentum $(\pi/2,\pi/2$)) for the $N=4$ (blue) and the $N=8$ (red) cylinder as a function of truncation threshold of the MPS.}
\label{fig:gap_hubbard}
\end{figure}

\par Now we apply our method to the two-dimensional Hubbard model on the square lattice \cite{Qin2021} at half filling, with Hamiltonian
\begin{multline} \label{eq:hubbard}
H_{\text{Hubbard}} = - \sum_{\sigma=\{\uparrow,\downarrow\}} \sum_{\braket{ij}} \left( c_{\sigma,i}\dag c_{\sigma,j}\pdag + c_{\sigma,j}\dag c_{\sigma,i}\pdag \right)  \\ + U \sum_i \left( c_{\uparrow,i}\dag c_{\uparrow,i}\pdag \right) \left( c_{\downarrow,i}\dag c_{\downarrow,i}\pdag \right),
\end{multline}
where $c_{\sigma,i}$ and $c_{\sigma,i}\dag$ are fermionic creation and annihilation operators with spin $\sigma$ at site $i$. In our MPS representation for the ground state on the cylinder, we fix the filling by implementing the $\U(1)$ symmetry for the electron charge, use $\SU(2)$ spin-rotation symmetry (which is left unbroken on a cylinder with finite circumference) and implement a graded $\Z_2$ symmetry \cite{Bultinck2017} for encorporating fermionic statistics. We work at $U=12$.
\par We optimize the ground state on infinite cylinders of circumference $N=4$ and $N=8$ using the vumps algorithm \cite{ZaunerStauber2018}, and look at the excitation spectrum on top of the ground state. Using the $\U(1)$ and $\SU(2)$ symmetries in the MPS representation, we can fix the charge and spin quantum numbers $(q_c,q_s)$ in the excited state ansatz. The ansatz will provide us with a variational energy for the lowest-lying state at a given momentum $(p_x,p_y)$ and quantum numbers $(q_c,q_s)$; this can be a multi-particle state that is obtained from combining different quasiparticles with other momenta and quantum numbers -- in the thermodynamic limit, the momenta, energies and quantum numbers of quasiparticles can simply be added (or fused, in case of an $\SU(2)$ quantum number). 
\par First, we study the magnon excitations with quantum numbers $(0,1)$ in the top panel of Fig.~\ref{fig:disp_hubbard}. We observe a strong minimum in the X-point at $(\pi,\pi)$, in agreement with the fact that the model effectively behaves as a Heisenberg antiferromagnet in the charge-neutral sector. We also see a minimum at $(0,0)$, corresponding to a two-magnon state. We observe that the  magnon gap at $(\pi,\pi)$ converges quickly with bond dimension (not shown), which points to the fact that the magnon is a well-defined quasiparticle for which our ansatz is ideally suited. Correspondingly, we observe that the energy of the two-magnon state at $(0,0)$ converges a lot slower (not shown).
\par Next, we consider the charged excitations with quantum numbers $(-1,\frac{1}{2})$. The hole dispersion can be directly observed in angle-resolved photoemission spectroscopy, and has gained a lot of (renewed) theoretical attention due to recent cold-atom experiments \cite{Chiu2019, Bohrdt2019}. In particular, the nature of the magnetic polaron as a bound state of chargon and spinon \cite{Beran1997, Laughlin1997, Bohrdt2020} has been investigated in some detail. In the bottom panel of Fig.~\ref{fig:disp_hubbard} we have plotted three momentum cuts for the $N=4$ cyinder, and we observe a minimum in the S point at $(\pi/2,\pi/2)$. This feature is in agreement with a recent numerical study of the $t$-$J$ model \cite{Bohrdt2020}. The $p_y=0$ and $p_y=\pi$ cuts show minima away from the edges of the Brillouin zone, and a level crossing at higher energies. Since the spin sector is gapless, low-energy two-particle excitations can be created by adding a magnon to a hole excitation; therefore, the dispersion we observe is the one of the lower edge of a multi-particle continuum, and we can not directly interpret this dispersion as a quasiparticle mode. Note that the hole excitation energies are negative for the Hamiltonian in Eq.~\ref{eq:hubbard}; adding a chemical potential of $\mu=U/2$ would stabilize the half-filled ground state and yield a particle-hole symmetric spectrum with a positive charge gap.
\par In Fig.~\ref{fig:gap_hubbard} we plot this minimum of the hole dispersion, i.e. the charge gap, as a function of Schmidt-value threshold\footnote{The Schmidt-value threshold is the approximate value of the smallest Schmidt value in all symmetry sectors of the MPS; it is the analog of the truncation threshold in (i)DMRG simulations, which does not have a clear meaning in variational optimization schemes without truncation steps.} in the MPS ground state for the $N=4$ and the $N=8$ cylinders. We observe that convergence is rather slow, which points to the composite nature of the excitation: The quasiparticle ansatz needs a large bond dimension for describing the extended bound-state complex. As Fig.~\ref{fig:gap_hubbard} shows, the charge gap is not completely converged for the $N=8$ cylinder with our highest bond dimension ($D\approx7000$), but presumably a larger-scale simulation could find a fully converged value also for the eight-site cylinder.

\section{Conclusions}

\noindent In this paper we have applied open-boundary MPS algorithms to systems with periodic boundary conditions. In particular, although translation symmetry is broken by the MPS representation, we have shown that momentum emerges as a good quantum number. This can be exploited for labeling excitations on top of MPS ground states. 
\par We have also introduced an ansatz for capturing excited states in finite systems with open boundary conditions, which is a very efficient method for computing e.g. the spectral gap, and which can be straightforwardly implemented on top of an existing DMRG or MPS ground-state code. We expect that this approach can be of great use to the DMRG/MPS community.
\par Our method was shown to work very well for obtaining CFT finite-size spectra of critical models. We expect this will prove useful for characterizing the effective CFT for 1-D quantum and 2-D classical models, as well as to compute entanglement spectra of projected entangled-pair states in order to characterize the (chiral) edge field theory \cite{Poilblanc2016}.
\par Applying the framework to 2-D cylinders has shown that we can obtain excitation spectra of challenging models such as spin liquids or (doped) Hubbard models. Indeed, whereas simulating time evolution is often prohibitively expensive for these models and excitation spectra are, therefore, rather infeasible to compute, the excitation ansatz can be applied with a cost that is similar to a ground-state simulation. In that respect, it would be extremely interesting to look at fractionalized excitations such as spinons or holons: In principle, the excitation ansatz can be straightforwardly generalized to also capture these topological excitations \cite{ZaunerStauber2018b, Vanderstraeten2020}. It would be interesting to, e.g., further investigate hole dynamics in Hubbard models with frustration; here, the excitation ansatz can be complemented with time-evolution approaches \cite{Kjall2011, Gohlke2017} to yield more insight into the composite nature of the hole. Another question is how this MPS excitation ansatz on the cylinder compares to the excitation ansatz for projected entangled-pair states \cite{Vanderstraeten2015b, Vanderstraeten2019b, Ponsioen2020}, which is formulated directly on the infinite plane.

\section{Acknowledgements}

\noindent The authors would like to thank Ji-Yao Chen, Boris Ponsioen, Philippe Corboz, Michael Knap and Frank Pollmann for inspiring discussions. This work was supported by the Research Foundation Flanders (G0E1520N, G0E1820N) and the ERC grants QUTE (647905) and ERQUAF (715861).

\bibliography{./bibliography}

\newpage
\appendix
\section{Pseudocode for excitation ansatz with open-boundary MPS}
\label{sec:appendix}

\noindent In this Appendix we elaborate on the implementation details of the quasiparticle ansatz for finite systems. In particular, we will provide all formulas and pseudocode for computing the action of the effective Hamiltonian that appears in the eigenvalue equation
\begin{equation} \label{eq:a1}
\sum_{j} (H_{\mathrm{eff}})_{ij} \vec{X}_j = \omega \vec{X}_i,
\end{equation}
which can be fed into an iterative eigensolver to find the first few low-lying excitations in the system.
\par A finite MPS can be gauged in such a way that every tensor left and right from a given site is respectively left/right isometric
\begin{equation}
\ket{\Psi(A_1\dots A_N} = \diagram{p1}{2}.
\end{equation}
In what follows we will work in the MPO-representation of the Hamiltonian \cite{McCulloch2007, Schollwock2011}. We will need the partially contracted environments of the state, much like in existing DMRG codes:
\begin{equation}
\diagram{a1}{2} = \diagram{a1}{1}, \qquad \diagram{a1}{5} = \diagram{a1}{4} .
\end{equation}
We will also require $V_i$, the null space of $A_i^l$, such that
\begin{equation}
\diagram{p1}{6} = 0, \qquad \diagram{p1}{7}=\diagram{p1}{8}.
\end{equation}
All degrees of freedom can then be absorbed in tensors $X_i$, as we have seen earlier
\begin{multline}
    \diagram{p1}{4} = \diagram{p1}{5},  \\
    \ket{\Phi_i(X_i)} = \diagram{p1}{9}.
\end{multline}
The last two quantities we need to define are $\sigma^{l}_i$ and $\sigma^{r}_i$, satisfying:
\begin{equation}
\diagram{a1}{7} = \diagram{a1}{5}+\diagram{a1}{6}
\label{eq:rhobl}
\end{equation}
and
\begin{equation}
\diagram{a1}{10} = \diagram{a1}{8}+\diagram{a1}{9} .
\label{eq:rhobr}
\end{equation}
All these quantities together make $H_{\mathrm{eff}}\vec{X}$ take on a pleasant form. For every site $i$, one has to calculate the following tensor contractions to get the derivative with respect to the tensor $B_i'$:
\begin{multline}
    \diagram{a1}{14} = \diagram{a1}{12} \\ + \diagram{a1}{11}+\diagram{a1}{13}.
    \label{eq:Bderiv}
\end{multline}
The fundamental degrees of freedom are the $X_i'$ tensors, which can then be extracted by projecting down on $V_i$:
\begin{equation}
    X'_i =\diagram{a1}{15}.
    \label{eq:projdown}
\end{equation}
This concludes the action of the effective Hamiltonian, which indeed appears as a linear operator on the set of tensors $X_i$. The full action of $H_{\mathrm{eff}}$ is, therefore, denoted as
\begin{equation}
\vec{X}_i' = \sum_{j} (H_{\mathrm{eff}})_{ij} \vec{X}_j,
\end{equation}
such that the above eigenvalue equation \ref{eq:a1} can be solved iteratively.

\par The pseudocode for the action of $H_{\mathrm{eff}}$ on an input list of tensors $X_i$ can be found in Algorithm \ref{alg:code}. Eigenvectors can be found using any iterative eigensolver. This algorithm is very simple, straightforward to implement and requires similar contractions as in `usual' DMRG codes. Not much changes when going to the thermodynamic limit, except the calculation for $\sigma^{l}_i$, $\sigma^{r}_i$ involves infinite sums \cite{Vanderstraeten2019}.

\begin{figure}[H]
	\begin{algorithm}[H]
		\begin{algorithmic}[1]
		    \State Inputs $(\vec{X},\vec{\rho^l},\vec{\rho^r})$
		    \State $\vec{X}' \gets 0$, $\sigma^{l}_i \gets 0$, $\sigma^{r}_i \gets 0$ \Comment{initialization}
		    \For{$i \in [1,len(\Vec{X})]$} \Comment{environments}
		        \State calculate $\sigma^{l}_{i+1}$ \Comment{Eq.~\ref{eq:rhobl}}
		        \State calculate $\sigma^{r}_{end-i}$ \Comment{Eq.~\ref{eq:rhobr}}
		    \EndFor
		    
		    \For{$i \in [1,len(\Vec{X})]$} 
		        \State calculate $T_i$ \Comment{Eq.~\ref{eq:Bderiv}}
		        \State calculate $X_i'$ \Comment{Eq.~\ref{eq:projdown}}
		    \EndFor
		    
		    \State \Return $\Vec{X}'$
		\end{algorithmic}
		\caption{Pseudocode for the action of $H_{\mathrm{eff}}(\Vec{X})$}
		\label{alg:code}
	\end{algorithm}
\end{figure}

\end{document}

%% file: preamble.tex
\usepackage{graphicx}
\usepackage{graphics}
\usepackage{amsmath}
\usepackage{amssymb}
\usepackage{amsfonts}
\usepackage{dsfont}
\usepackage{braket}
\usepackage{color}
\usepackage{braket,slashed}
\usepackage[mathscr]{euscript}
\definecolor{darkblue}{rgb}{0, 0, 0.8}
\usepackage[colorlinks=true, breaklinks=true, linkcolor=red, citecolor=blue, urlcolor=blue]{hyperref} 
\usepackage{hyperref}
\usepackage{subfigure}
\usepackage{xfrac}
\usepackage{bm}
\usepackage{kantlipsum}
\usepackage{enumitem}
\usepackage{tikz}
\usepackage{framed}
\usepackage{graphicx}
\usepackage{subfigure}
\usepackage{algorithm}
\usepackage[noend]{algpseudocode}

\allowdisplaybreaks[1]

\newcommand{\code}[1]{\texttt{#1}}

\renewcommand{\dag}{^\dagger}
\newcommand{\pdag}{^{\phantom{\dagger}}}

\newcommand{\e}{\ensuremath{\mathrm{e}}}

\newcommand{\SU}{\ensuremath{\mathrm{SU}}}
\newcommand{\Z}{\ensuremath{\mathbb{Z}}}
\newcommand{\U}{\ensuremath{\mathrm{U}}}

\newcommand{\diagram}[2]{\;\vcenter{\hbox{\includegraphics[scale=0.32,page=#2]{./#1.pdf}}}\;}

%% file: main.bbl
\begin{thebibliography}{63}%
\makeatletter
\providecommand \@ifxundefined [1]{%
 \@ifx{#1\undefined}
}%
\providecommand \@ifnum [1]{%
 \ifnum #1\expandafter \@firstoftwo
 \else \expandafter \@secondoftwo
 \fi
}%
\providecommand \@ifx [1]{%
 \ifx #1\expandafter \@firstoftwo
 \else \expandafter \@secondoftwo
 \fi
}%
\providecommand \natexlab [1]{#1}%
\providecommand \enquote  [1]{``#1''}%
\providecommand \bibnamefont  [1]{#1}%
\providecommand \bibfnamefont [1]{#1}%
\providecommand \citenamefont [1]{#1}%
\providecommand \href@noop [0]{\@secondoftwo}%
\providecommand \href [0]{\begingroup \@sanitize@url \@href}%
\providecommand \@href[1]{\@@startlink{#1}\@@href}%
\providecommand \@@href[1]{\endgroup#1\@@endlink}%
\providecommand \@sanitize@url [0]{\catcode `\\12\catcode `\$12\catcode
  `\&12\catcode `\#12\catcode `\^12\catcode `\_12\catcode `\%12\relax}%
\providecommand \@@startlink[1]{}%
\providecommand \@@endlink[0]{}%
\providecommand \url  [0]{\begingroup\@sanitize@url \@url }%
\providecommand \@url [1]{\endgroup\@href {#1}{\urlprefix }}%
\providecommand \urlprefix  [0]{URL }%
\providecommand \Eprint [0]{\href }%
\providecommand \doibase [0]{https://doi.org/}%
\providecommand \selectlanguage [0]{\@gobble}%
\providecommand \bibinfo  [0]{\@secondoftwo}%
\providecommand \bibfield  [0]{\@secondoftwo}%
\providecommand \translation [1]{[#1]}%
\providecommand \BibitemOpen [0]{}%
\providecommand \bibitemStop [0]{}%
\providecommand \bibitemNoStop [0]{.\EOS\space}%
\providecommand \EOS [0]{\spacefactor3000\relax}%
\providecommand \BibitemShut  [1]{\csname bibitem#1\endcsname}%
\let\auto@bib@innerbib\@empty
\bibitem [{\citenamefont {Schollw\"ock}(2011)}]{Schollwock2011}%
  \BibitemOpen
  \bibfield  {author} {\bibinfo {author} {\bibfnamefont {U.}~\bibnamefont
  {Schollw\"ock}},\ }\bibfield  {title} {\bibinfo {title} {The density-matrix
  renormalization group in the age of matrix product states},\ }\href
  {https://doi.org/https://doi.org/10.1016/j.aop.2010.09.012} {\bibfield
  {journal} {\bibinfo  {journal} {Annals of Physics}\ }\textbf {\bibinfo
  {volume} {326}},\ \bibinfo {pages} {96 } (\bibinfo {year}
  {2011})}\BibitemShut {NoStop}%
\bibitem [{\citenamefont {Cirac}\ \emph {et~al.}(2020)\citenamefont {Cirac},
  \citenamefont {Perez-Garcia}, \citenamefont {Schuch},\ and\ \citenamefont
  {Verstraete}}]{Cirac2020}%
  \BibitemOpen
  \bibfield  {author} {\bibinfo {author} {\bibfnamefont {J.~I.}\ \bibnamefont
  {Cirac}}, \bibinfo {author} {\bibfnamefont {D.}~\bibnamefont {Perez-Garcia}},
  \bibinfo {author} {\bibfnamefont {N.}~\bibnamefont {Schuch}},\ and\ \bibinfo
  {author} {\bibfnamefont {F.}~\bibnamefont {Verstraete}},\ }\bibfield  {title}
  {\bibinfo {title} {Matrix product states and projected entangled pair states:
  Concepts, symmetries, and theorems},\ }\href
  {https://arxiv.org/abs/2011.12127} {\bibfield  {journal} {\bibinfo  {journal}
  {arXiv:2011.12127}\ } (\bibinfo {year} {2020})}\BibitemShut {NoStop}%
\bibitem [{\citenamefont {White}(1992)}]{White1992}%
  \BibitemOpen
  \bibfield  {author} {\bibinfo {author} {\bibfnamefont {S.~R.}\ \bibnamefont
  {White}},\ }\bibfield  {title} {\bibinfo {title} {Density matrix formulation
  for quantum renormalization groups},\ }\href
  {https://doi.org/10.1103/PhysRevLett.69.2863} {\bibfield  {journal} {\bibinfo
   {journal} {Phys. Rev. Lett.}\ }\textbf {\bibinfo {volume} {69}},\ \bibinfo
  {pages} {2863} (\bibinfo {year} {1992})}\BibitemShut {NoStop}%
\bibitem [{\citenamefont {White}(1993)}]{White1993}%
  \BibitemOpen
  \bibfield  {author} {\bibinfo {author} {\bibfnamefont {S.~R.}\ \bibnamefont
  {White}},\ }\bibfield  {title} {\bibinfo {title} {Density-matrix algorithms
  for quantum renormalization groups},\ }\href
  {https://doi.org/10.1103/PhysRevB.48.10345} {\bibfield  {journal} {\bibinfo
  {journal} {Phys. Rev. B}\ }\textbf {\bibinfo {volume} {48}},\ \bibinfo
  {pages} {10345} (\bibinfo {year} {1993})}\BibitemShut {NoStop}%
\bibitem [{\citenamefont {Rommer}\ and\ \citenamefont
  {\"Ostlund}(1997)}]{Rommer1997}%
  \BibitemOpen
  \bibfield  {author} {\bibinfo {author} {\bibfnamefont {S.}~\bibnamefont
  {Rommer}}\ and\ \bibinfo {author} {\bibfnamefont {S.}~\bibnamefont
  {\"Ostlund}},\ }\bibfield  {title} {\bibinfo {title} {Class of ansatz wave
  functions for one-dimensional spin systems and their relation to the density
  matrix renormalization group},\ }\href
  {https://doi.org/10.1103/PhysRevB.55.2164} {\bibfield  {journal} {\bibinfo
  {journal} {Phys. Rev. B}\ }\textbf {\bibinfo {volume} {55}},\ \bibinfo
  {pages} {2164} (\bibinfo {year} {1997})}\BibitemShut {NoStop}%
\bibitem [{\citenamefont {Verstraete}\ \emph {et~al.}(2004)\citenamefont
  {Verstraete}, \citenamefont {Porras},\ and\ \citenamefont
  {Cirac}}]{Verstraete2004}%
  \BibitemOpen
  \bibfield  {author} {\bibinfo {author} {\bibfnamefont {F.}~\bibnamefont
  {Verstraete}}, \bibinfo {author} {\bibfnamefont {D.}~\bibnamefont {Porras}},\
  and\ \bibinfo {author} {\bibfnamefont {J.~I.}\ \bibnamefont {Cirac}},\
  }\bibfield  {title} {\bibinfo {title} {Density matrix renormalization group
  and periodic boundary conditions: A quantum information perspective},\ }\href
  {https://doi.org/10.1103/PhysRevLett.93.227205} {\bibfield  {journal}
  {\bibinfo  {journal} {Phys. Rev. Lett.}\ }\textbf {\bibinfo {volume} {93}},\
  \bibinfo {pages} {227205} (\bibinfo {year} {2004})}\BibitemShut {NoStop}%
\bibitem [{\citenamefont {Perez-Garcia}\ \emph {et~al.}(2006)\citenamefont
  {Perez-Garcia}, \citenamefont {Verstraete}, \citenamefont {Wolf},\ and\
  \citenamefont {Cirac}}]{PerezGarcia2006}%
  \BibitemOpen
  \bibfield  {author} {\bibinfo {author} {\bibfnamefont {D.}~\bibnamefont
  {Perez-Garcia}}, \bibinfo {author} {\bibfnamefont {F.}~\bibnamefont
  {Verstraete}}, \bibinfo {author} {\bibfnamefont {M.~M.}\ \bibnamefont
  {Wolf}},\ and\ \bibinfo {author} {\bibfnamefont {J.~I.}\ \bibnamefont
  {Cirac}},\ }\bibfield  {title} {\bibinfo {title} {Matrix product state
  representations},\ }\href {https://arxiv.org/abs/quant-ph/0608197} {\bibfield
   {journal} {\bibinfo  {journal} {arXiv:quant-ph/0608197}\ } (\bibinfo {year}
  {2006})}\BibitemShut {NoStop}%
\bibitem [{\citenamefont {\"Ostlund}\ and\ \citenamefont
  {Rommer}(1995)}]{Ostlund1995}%
  \BibitemOpen
  \bibfield  {author} {\bibinfo {author} {\bibfnamefont {S.}~\bibnamefont
  {\"Ostlund}}\ and\ \bibinfo {author} {\bibfnamefont {S.}~\bibnamefont
  {Rommer}},\ }\bibfield  {title} {\bibinfo {title} {Thermodynamic limit of
  density matrix renormalization},\ }\href
  {https://doi.org/10.1103/PhysRevLett.75.3537} {\bibfield  {journal} {\bibinfo
   {journal} {Phys. Rev. Lett.}\ }\textbf {\bibinfo {volume} {75}},\ \bibinfo
  {pages} {3537} (\bibinfo {year} {1995})}\BibitemShut {NoStop}%
\bibitem [{\citenamefont {Vidal}(2007)}]{Vidal2007}%
  \BibitemOpen
  \bibfield  {author} {\bibinfo {author} {\bibfnamefont {G.}~\bibnamefont
  {Vidal}},\ }\bibfield  {title} {\bibinfo {title} {Classical simulation of
  infinite-size quantum lattice systems in one spatial dimension},\ }\href
  {https://doi.org/10.1103/PhysRevLett.98.070201} {\bibfield  {journal}
  {\bibinfo  {journal} {Phys. Rev. Lett.}\ }\textbf {\bibinfo {volume} {98}},\
  \bibinfo {pages} {070201} (\bibinfo {year} {2007})}\BibitemShut {NoStop}%
\bibitem [{\citenamefont {McCulloch}(2008)}]{McCulloch2008}%
  \BibitemOpen
  \bibfield  {author} {\bibinfo {author} {\bibfnamefont {I.~P.}\ \bibnamefont
  {McCulloch}},\ }\bibfield  {title} {\bibinfo {title} {Infinite size density
  matrix renormalization group, revisited},\ }\href
  {https://arxiv.org/abs/0804.2509} {\bibfield  {journal} {\bibinfo  {journal}
  {arXiv:0804.2509}\ } (\bibinfo {year} {2008})}\BibitemShut {NoStop}%
\bibitem [{\citenamefont {Haegeman}\ \emph {et~al.}(2011)\citenamefont
  {Haegeman}, \citenamefont {Cirac}, \citenamefont {Osborne}, \citenamefont
  {Pi\ifmmode~\check{z}\else \v{z}\fi{}orn}, \citenamefont {Verschelde},\ and\
  \citenamefont {Verstraete}}]{Haegeman2011}%
  \BibitemOpen
  \bibfield  {author} {\bibinfo {author} {\bibfnamefont {J.}~\bibnamefont
  {Haegeman}}, \bibinfo {author} {\bibfnamefont {J.~I.}\ \bibnamefont {Cirac}},
  \bibinfo {author} {\bibfnamefont {T.~J.}\ \bibnamefont {Osborne}}, \bibinfo
  {author} {\bibfnamefont {I.}~\bibnamefont {Pi\ifmmode~\check{z}\else
  \v{z}\fi{}orn}}, \bibinfo {author} {\bibfnamefont {H.}~\bibnamefont
  {Verschelde}},\ and\ \bibinfo {author} {\bibfnamefont {F.}~\bibnamefont
  {Verstraete}},\ }\bibfield  {title} {\bibinfo {title} {Time-dependent
  variational principle for quantum lattices},\ }\href
  {https://doi.org/10.1103/PhysRevLett.107.070601} {\bibfield  {journal}
  {\bibinfo  {journal} {Phys. Rev. Lett.}\ }\textbf {\bibinfo {volume} {107}},\
  \bibinfo {pages} {070601} (\bibinfo {year} {2011})}\BibitemShut {NoStop}%
\bibitem [{\citenamefont {Vanderstraeten}\ \emph
  {et~al.}(2019{\natexlab{a}})\citenamefont {Vanderstraeten}, \citenamefont
  {Haegeman},\ and\ \citenamefont {Verstraete}}]{Vanderstraeten2019}%
  \BibitemOpen
  \bibfield  {author} {\bibinfo {author} {\bibfnamefont {L.}~\bibnamefont
  {Vanderstraeten}}, \bibinfo {author} {\bibfnamefont {J.}~\bibnamefont
  {Haegeman}},\ and\ \bibinfo {author} {\bibfnamefont {F.}~\bibnamefont
  {Verstraete}},\ }\bibfield  {title} {\bibinfo {title} {{Tangent-space methods
  for uniform matrix product states}},\ }\href
  {https://doi.org/10.21468/SciPostPhysLectNotes.7} {\bibfield  {journal}
  {\bibinfo  {journal} {SciPost Phys. Lect. Notes}\ ,\ \bibinfo {pages} {7}}
  (\bibinfo {year} {2019}{\natexlab{a}})}\BibitemShut {NoStop}%
\bibitem [{\citenamefont {Haegeman}\ \emph {et~al.}(2012)\citenamefont
  {Haegeman}, \citenamefont {Pirvu}, \citenamefont {Weir}, \citenamefont
  {Cirac}, \citenamefont {Osborne}, \citenamefont {Verschelde},\ and\
  \citenamefont {Verstraete}}]{Haegeman2012}%
  \BibitemOpen
  \bibfield  {author} {\bibinfo {author} {\bibfnamefont {J.}~\bibnamefont
  {Haegeman}}, \bibinfo {author} {\bibfnamefont {B.}~\bibnamefont {Pirvu}},
  \bibinfo {author} {\bibfnamefont {D.~J.}\ \bibnamefont {Weir}}, \bibinfo
  {author} {\bibfnamefont {J.~I.}\ \bibnamefont {Cirac}}, \bibinfo {author}
  {\bibfnamefont {T.~J.}\ \bibnamefont {Osborne}}, \bibinfo {author}
  {\bibfnamefont {H.}~\bibnamefont {Verschelde}},\ and\ \bibinfo {author}
  {\bibfnamefont {F.}~\bibnamefont {Verstraete}},\ }\bibfield  {title}
  {\bibinfo {title} {Variational matrix product ansatz for dispersion
  relations},\ }\href {https://doi.org/10.1103/PhysRevB.85.100408} {\bibfield
  {journal} {\bibinfo  {journal} {Phys. Rev. B}\ }\textbf {\bibinfo {volume}
  {85}},\ \bibinfo {pages} {100408} (\bibinfo {year} {2012})}\BibitemShut
  {NoStop}%
\bibitem [{\citenamefont {Feynman}(1953)}]{Feynman1953}%
  \BibitemOpen
  \bibfield  {author} {\bibinfo {author} {\bibfnamefont {R.~P.}\ \bibnamefont
  {Feynman}},\ }\bibfield  {title} {\bibinfo {title} {Atomic theory of the
  $\ensuremath{\lambda}$ transition in helium},\ }\href
  {https://doi.org/10.1103/PhysRev.91.1291} {\bibfield  {journal} {\bibinfo
  {journal} {Phys. Rev.}\ }\textbf {\bibinfo {volume} {91}},\ \bibinfo {pages}
  {1291} (\bibinfo {year} {1953})}\BibitemShut {NoStop}%
\bibitem [{\citenamefont {Girvin}\ \emph {et~al.}(1985)\citenamefont {Girvin},
  \citenamefont {MacDonald},\ and\ \citenamefont {Platzman}}]{Girvin1985}%
  \BibitemOpen
  \bibfield  {author} {\bibinfo {author} {\bibfnamefont {S.~M.}\ \bibnamefont
  {Girvin}}, \bibinfo {author} {\bibfnamefont {A.~H.}\ \bibnamefont
  {MacDonald}},\ and\ \bibinfo {author} {\bibfnamefont {P.~M.}\ \bibnamefont
  {Platzman}},\ }\bibfield  {title} {\bibinfo {title} {Collective-excitation
  gap in the fractional quantum hall effect},\ }\href
  {https://doi.org/10.1103/PhysRevLett.54.581} {\bibfield  {journal} {\bibinfo
  {journal} {Phys. Rev. Lett.}\ }\textbf {\bibinfo {volume} {54}},\ \bibinfo
  {pages} {581} (\bibinfo {year} {1985})}\BibitemShut {NoStop}%
\bibitem [{\citenamefont {Arovas}\ \emph {et~al.}(1988)\citenamefont {Arovas},
  \citenamefont {Auerbach},\ and\ \citenamefont {Haldane}}]{Arovas1988}%
  \BibitemOpen
  \bibfield  {author} {\bibinfo {author} {\bibfnamefont {D.~P.}\ \bibnamefont
  {Arovas}}, \bibinfo {author} {\bibfnamefont {A.}~\bibnamefont {Auerbach}},\
  and\ \bibinfo {author} {\bibfnamefont {F.~D.~M.}\ \bibnamefont {Haldane}},\
  }\bibfield  {title} {\bibinfo {title} {Extended heisenberg models of
  antiferromagnetism: Analogies to the fractional quantum hall effect},\ }\href
  {https://doi.org/10.1103/PhysRevLett.60.531} {\bibfield  {journal} {\bibinfo
  {journal} {Phys. Rev. Lett.}\ }\textbf {\bibinfo {volume} {60}},\ \bibinfo
  {pages} {531} (\bibinfo {year} {1988})}\BibitemShut {NoStop}%
\bibitem [{\citenamefont {Takahashi}(1994)}]{Takahashi1994}%
  \BibitemOpen
  \bibfield  {author} {\bibinfo {author} {\bibfnamefont {M.}~\bibnamefont
  {Takahashi}},\ }\bibfield  {title} {\bibinfo {title} {Excitation spectra of
  s=1 antiferromagnetic chains},\ }\href
  {https://doi.org/10.1103/PhysRevB.50.3045} {\bibfield  {journal} {\bibinfo
  {journal} {Phys. Rev. B}\ }\textbf {\bibinfo {volume} {50}},\ \bibinfo
  {pages} {3045} (\bibinfo {year} {1994})}\BibitemShut {NoStop}%
\bibitem [{\citenamefont {S\o{}rensen}\ and\ \citenamefont
  {Affleck}(1994)}]{Sorensen1994}%
  \BibitemOpen
  \bibfield  {author} {\bibinfo {author} {\bibfnamefont {E.~S.}\ \bibnamefont
  {S\o{}rensen}}\ and\ \bibinfo {author} {\bibfnamefont {I.}~\bibnamefont
  {Affleck}},\ }\bibfield  {title} {\bibinfo {title} {Equal-time correlations
  in haldane-gap antiferromagnets},\ }\href
  {https://doi.org/10.1103/PhysRevB.49.15771} {\bibfield  {journal} {\bibinfo
  {journal} {Phys. Rev. B}\ }\textbf {\bibinfo {volume} {49}},\ \bibinfo
  {pages} {15771} (\bibinfo {year} {1994})}\BibitemShut {NoStop}%
\bibitem [{\citenamefont {Bera}\ \emph {et~al.}(2017)\citenamefont {Bera},
  \citenamefont {Lake}, \citenamefont {Essler}, \citenamefont {Vanderstraeten},
  \citenamefont {Hubig}, \citenamefont {Schollw\"ock}, \citenamefont {Islam},
  \citenamefont {Schneidewind},\ and\ \citenamefont
  {Quintero-Castro}}]{Bera2017}%
  \BibitemOpen
  \bibfield  {author} {\bibinfo {author} {\bibfnamefont {A.~K.}\ \bibnamefont
  {Bera}}, \bibinfo {author} {\bibfnamefont {B.}~\bibnamefont {Lake}}, \bibinfo
  {author} {\bibfnamefont {F.~H.~L.}\ \bibnamefont {Essler}}, \bibinfo {author}
  {\bibfnamefont {L.}~\bibnamefont {Vanderstraeten}}, \bibinfo {author}
  {\bibfnamefont {C.}~\bibnamefont {Hubig}}, \bibinfo {author} {\bibfnamefont
  {U.}~\bibnamefont {Schollw\"ock}}, \bibinfo {author} {\bibfnamefont {A.~T.
  M.~N.}\ \bibnamefont {Islam}}, \bibinfo {author} {\bibfnamefont
  {A.}~\bibnamefont {Schneidewind}},\ and\ \bibinfo {author} {\bibfnamefont
  {D.~L.}\ \bibnamefont {Quintero-Castro}},\ }\bibfield  {title} {\bibinfo
  {title} {Spinon confinement in a quasi-one-dimensional anisotropic heisenberg
  magnet},\ }\href {https://doi.org/10.1103/PhysRevB.96.054423} {\bibfield
  {journal} {\bibinfo  {journal} {Phys. Rev. B}\ }\textbf {\bibinfo {volume}
  {96}},\ \bibinfo {pages} {054423} (\bibinfo {year} {2017})}\BibitemShut
  {NoStop}%
\bibitem [{\citenamefont {Zauner-Stauber}\ \emph
  {et~al.}(2018{\natexlab{a}})\citenamefont {Zauner-Stauber}, \citenamefont
  {Vanderstraeten}, \citenamefont {Haegeman}, \citenamefont {McCulloch},\ and\
  \citenamefont {Verstraete}}]{ZaunerStauber2018b}%
  \BibitemOpen
  \bibfield  {author} {\bibinfo {author} {\bibfnamefont {V.}~\bibnamefont
  {Zauner-Stauber}}, \bibinfo {author} {\bibfnamefont {L.}~\bibnamefont
  {Vanderstraeten}}, \bibinfo {author} {\bibfnamefont {J.}~\bibnamefont
  {Haegeman}}, \bibinfo {author} {\bibfnamefont {I.~P.}\ \bibnamefont
  {McCulloch}},\ and\ \bibinfo {author} {\bibfnamefont {F.}~\bibnamefont
  {Verstraete}},\ }\bibfield  {title} {\bibinfo {title} {Topological nature of
  spinons and holons: Elementary excitations from matrix product states with
  conserved symmetries},\ }\href {https://doi.org/10.1103/PhysRevB.97.235155}
  {\bibfield  {journal} {\bibinfo  {journal} {Phys. Rev. B}\ }\textbf {\bibinfo
  {volume} {97}},\ \bibinfo {pages} {235155} (\bibinfo {year}
  {2018}{\natexlab{a}})}\BibitemShut {NoStop}%
\bibitem [{\citenamefont {Vanderstraeten}\ \emph {et~al.}(2018)\citenamefont
  {Vanderstraeten}, \citenamefont {{Van Damme}}, \citenamefont {B\"uchler},\
  and\ \citenamefont {Verstraete}}]{Vanderstraeten2018}%
  \BibitemOpen
  \bibfield  {author} {\bibinfo {author} {\bibfnamefont {L.}~\bibnamefont
  {Vanderstraeten}}, \bibinfo {author} {\bibfnamefont {M.}~\bibnamefont {{Van
  Damme}}}, \bibinfo {author} {\bibfnamefont {H.~P.}\ \bibnamefont
  {B\"uchler}},\ and\ \bibinfo {author} {\bibfnamefont {F.}~\bibnamefont
  {Verstraete}},\ }\bibfield  {title} {\bibinfo {title} {Quasiparticles in
  quantum spin chains with long-range interactions},\ }\href
  {https://doi.org/10.1103/PhysRevLett.121.090603} {\bibfield  {journal}
  {\bibinfo  {journal} {Phys. Rev. Lett.}\ }\textbf {\bibinfo {volume} {121}},\
  \bibinfo {pages} {090603} (\bibinfo {year} {2018})}\BibitemShut {NoStop}%
\bibitem [{\citenamefont {Milsted}\ \emph {et~al.}(2013)\citenamefont
  {Milsted}, \citenamefont {Haegeman}, \citenamefont {Osborne},\ and\
  \citenamefont {Verstraete}}]{Milsted2013}%
  \BibitemOpen
  \bibfield  {author} {\bibinfo {author} {\bibfnamefont {A.}~\bibnamefont
  {Milsted}}, \bibinfo {author} {\bibfnamefont {J.}~\bibnamefont {Haegeman}},
  \bibinfo {author} {\bibfnamefont {T.~J.}\ \bibnamefont {Osborne}},\ and\
  \bibinfo {author} {\bibfnamefont {F.}~\bibnamefont {Verstraete}},\ }\bibfield
   {title} {\bibinfo {title} {Variational matrix product ansatz for nonuniform
  dynamics in the thermodynamic limit},\ }\href
  {https://doi.org/10.1103/PhysRevB.88.155116} {\bibfield  {journal} {\bibinfo
  {journal} {Phys. Rev. B}\ }\textbf {\bibinfo {volume} {88}},\ \bibinfo
  {pages} {155116} (\bibinfo {year} {2013})}\BibitemShut {NoStop}%
\bibitem [{\citenamefont {Phien}\ \emph {et~al.}(2012)\citenamefont {Phien},
  \citenamefont {Vidal},\ and\ \citenamefont {McCulloch}}]{Phien2012}%
  \BibitemOpen
  \bibfield  {author} {\bibinfo {author} {\bibfnamefont {H.~N.}\ \bibnamefont
  {Phien}}, \bibinfo {author} {\bibfnamefont {G.}~\bibnamefont {Vidal}},\ and\
  \bibinfo {author} {\bibfnamefont {I.~P.}\ \bibnamefont {McCulloch}},\
  }\bibfield  {title} {\bibinfo {title} {Infinite boundary conditions for
  matrix product state calculations},\ }\href
  {https://doi.org/10.1103/PhysRevB.86.245107} {\bibfield  {journal} {\bibinfo
  {journal} {Phys. Rev. B}\ }\textbf {\bibinfo {volume} {86}},\ \bibinfo
  {pages} {245107} (\bibinfo {year} {2012})}\BibitemShut {NoStop}%
\bibitem [{\citenamefont {Zauner}\ \emph {et~al.}(2015)\citenamefont {Zauner},
  \citenamefont {Ganahl}, \citenamefont {Evertz},\ and\ \citenamefont
  {Nishino}}]{Zauner2015}%
  \BibitemOpen
  \bibfield  {author} {\bibinfo {author} {\bibfnamefont {V.}~\bibnamefont
  {Zauner}}, \bibinfo {author} {\bibfnamefont {M.}~\bibnamefont {Ganahl}},
  \bibinfo {author} {\bibfnamefont {H.~G.}\ \bibnamefont {Evertz}},\ and\
  \bibinfo {author} {\bibfnamefont {T.}~\bibnamefont {Nishino}},\ }\bibfield
  {title} {\bibinfo {title} {Time evolution within a comoving window: scaling
  of signal fronts and magnetization plateaus after a local quench in quantum
  spin chains},\ }\href {https://doi.org/10.1088/0953-8984/27/42/425602}
  {\bibfield  {journal} {\bibinfo  {journal} {Journal of Physics: Condensed
  Matter}\ }\textbf {\bibinfo {volume} {27}},\ \bibinfo {pages} {425602}
  (\bibinfo {year} {2015})}\BibitemShut {NoStop}%
\bibitem [{\citenamefont {Kj\"all}\ \emph {et~al.}(2011)\citenamefont
  {Kj\"all}, \citenamefont {Pollmann},\ and\ \citenamefont
  {Moore}}]{Kjall2011}%
  \BibitemOpen
  \bibfield  {author} {\bibinfo {author} {\bibfnamefont {J.~A.}\ \bibnamefont
  {Kj\"all}}, \bibinfo {author} {\bibfnamefont {F.}~\bibnamefont {Pollmann}},\
  and\ \bibinfo {author} {\bibfnamefont {J.~E.}\ \bibnamefont {Moore}},\
  }\bibfield  {title} {\bibinfo {title} {Bound states and ${E}_{8}$ symmetry
  effects in perturbed quantum ising chains},\ }\href
  {https://doi.org/10.1103/PhysRevB.83.020407} {\bibfield  {journal} {\bibinfo
  {journal} {Phys. Rev. B}\ }\textbf {\bibinfo {volume} {83}},\ \bibinfo
  {pages} {020407} (\bibinfo {year} {2011})}\BibitemShut {NoStop}%
\bibitem [{\citenamefont {Gohlke}\ \emph {et~al.}(2017)\citenamefont {Gohlke},
  \citenamefont {Verresen}, \citenamefont {Moessner},\ and\ \citenamefont
  {Pollmann}}]{Gohlke2017}%
  \BibitemOpen
  \bibfield  {author} {\bibinfo {author} {\bibfnamefont {M.}~\bibnamefont
  {Gohlke}}, \bibinfo {author} {\bibfnamefont {R.}~\bibnamefont {Verresen}},
  \bibinfo {author} {\bibfnamefont {R.}~\bibnamefont {Moessner}},\ and\
  \bibinfo {author} {\bibfnamefont {F.}~\bibnamefont {Pollmann}},\ }\bibfield
  {title} {\bibinfo {title} {Dynamics of the kitaev-heisenberg model},\ }\href
  {https://doi.org/10.1103/PhysRevLett.119.157203} {\bibfield  {journal}
  {\bibinfo  {journal} {Phys. Rev. Lett.}\ }\textbf {\bibinfo {volume} {119}},\
  \bibinfo {pages} {157203} (\bibinfo {year} {2017})}\BibitemShut {NoStop}%
\bibitem [{\citenamefont {Francesco}\ \emph {et~al.}(2012)\citenamefont
  {Francesco}, \citenamefont {Mathieu},\ and\ \citenamefont
  {S{\'e}n{\'e}chal}}]{Francesco2012}%
  \BibitemOpen
  \bibfield  {author} {\bibinfo {author} {\bibfnamefont {P.}~\bibnamefont
  {Francesco}}, \bibinfo {author} {\bibfnamefont {P.}~\bibnamefont {Mathieu}},\
  and\ \bibinfo {author} {\bibfnamefont {D.}~\bibnamefont {S{\'e}n{\'e}chal}},\
  }\href@noop {} {\emph {\bibinfo {title} {Conformal field theory}}}\ (\bibinfo
   {publisher} {Springer Science \& Business Media},\ \bibinfo {year}
  {2012})\BibitemShut {NoStop}%
\bibitem [{\citenamefont {Porras}\ \emph {et~al.}(2006)\citenamefont {Porras},
  \citenamefont {Verstraete},\ and\ \citenamefont {Cirac}}]{Porras2006}%
  \BibitemOpen
  \bibfield  {author} {\bibinfo {author} {\bibfnamefont {D.}~\bibnamefont
  {Porras}}, \bibinfo {author} {\bibfnamefont {F.}~\bibnamefont {Verstraete}},\
  and\ \bibinfo {author} {\bibfnamefont {J.~I.}\ \bibnamefont {Cirac}},\
  }\bibfield  {title} {\bibinfo {title} {Renormalization algorithm for the
  calculation of spectra of interacting quantum systems},\ }\href
  {https://doi.org/10.1103/PhysRevB.73.014410} {\bibfield  {journal} {\bibinfo
  {journal} {Phys. Rev. B}\ }\textbf {\bibinfo {volume} {73}},\ \bibinfo
  {pages} {014410} (\bibinfo {year} {2006})}\BibitemShut {NoStop}%
\bibitem [{\citenamefont {Pirvu}\ \emph {et~al.}(2011)\citenamefont {Pirvu},
  \citenamefont {Verstraete},\ and\ \citenamefont {Vidal}}]{Pirvu2011}%
  \BibitemOpen
  \bibfield  {author} {\bibinfo {author} {\bibfnamefont {B.}~\bibnamefont
  {Pirvu}}, \bibinfo {author} {\bibfnamefont {F.}~\bibnamefont {Verstraete}},\
  and\ \bibinfo {author} {\bibfnamefont {G.}~\bibnamefont {Vidal}},\ }\bibfield
   {title} {\bibinfo {title} {Exploiting translational invariance in matrix
  product state simulations of spin chains with periodic boundary conditions},\
  }\href {https://doi.org/10.1103/PhysRevB.83.125104} {\bibfield  {journal}
  {\bibinfo  {journal} {Phys. Rev. B}\ }\textbf {\bibinfo {volume} {83}},\
  \bibinfo {pages} {125104} (\bibinfo {year} {2011})}\BibitemShut {NoStop}%
\bibitem [{\citenamefont {Pirvu}\ \emph
  {et~al.}(2012{\natexlab{a}})\citenamefont {Pirvu}, \citenamefont {Vidal},
  \citenamefont {Verstraete},\ and\ \citenamefont {Tagliacozzo}}]{Pirvu2012b}%
  \BibitemOpen
  \bibfield  {author} {\bibinfo {author} {\bibfnamefont {B.}~\bibnamefont
  {Pirvu}}, \bibinfo {author} {\bibfnamefont {G.}~\bibnamefont {Vidal}},
  \bibinfo {author} {\bibfnamefont {F.}~\bibnamefont {Verstraete}},\ and\
  \bibinfo {author} {\bibfnamefont {L.}~\bibnamefont {Tagliacozzo}},\
  }\bibfield  {title} {\bibinfo {title} {Matrix product states for critical
  spin chains: Finite-size versus finite-entanglement scaling},\ }\href
  {https://doi.org/10.1103/PhysRevB.86.075117} {\bibfield  {journal} {\bibinfo
  {journal} {Phys. Rev. B}\ }\textbf {\bibinfo {volume} {86}},\ \bibinfo
  {pages} {075117} (\bibinfo {year} {2012}{\natexlab{a}})}\BibitemShut
  {NoStop}%
\bibitem [{\citenamefont {Draxler}\ \emph {et~al.}(2017)\citenamefont
  {Draxler}, \citenamefont {Haegeman}, \citenamefont {Verstraete},\ and\
  \citenamefont {Rizzi}}]{Draxler2017}%
  \BibitemOpen
  \bibfield  {author} {\bibinfo {author} {\bibfnamefont {D.}~\bibnamefont
  {Draxler}}, \bibinfo {author} {\bibfnamefont {J.}~\bibnamefont {Haegeman}},
  \bibinfo {author} {\bibfnamefont {F.}~\bibnamefont {Verstraete}},\ and\
  \bibinfo {author} {\bibfnamefont {M.}~\bibnamefont {Rizzi}},\ }\bibfield
  {title} {\bibinfo {title} {Continuous matrix product states with periodic
  boundary conditions and an application to atomtronics},\ }\href
  {https://doi.org/10.1103/PhysRevB.95.045145} {\bibfield  {journal} {\bibinfo
  {journal} {Phys. Rev. B}\ }\textbf {\bibinfo {volume} {95}},\ \bibinfo
  {pages} {045145} (\bibinfo {year} {2017})}\BibitemShut {NoStop}%
\bibitem [{\citenamefont {Zou}\ \emph {et~al.}(2018)\citenamefont {Zou},
  \citenamefont {Milsted},\ and\ \citenamefont {Vidal}}]{Zou2018}%
  \BibitemOpen
  \bibfield  {author} {\bibinfo {author} {\bibfnamefont {Y.}~\bibnamefont
  {Zou}}, \bibinfo {author} {\bibfnamefont {A.}~\bibnamefont {Milsted}},\ and\
  \bibinfo {author} {\bibfnamefont {G.}~\bibnamefont {Vidal}},\ }\bibfield
  {title} {\bibinfo {title} {Conformal data and renormalization group flow in
  critical quantum spin chains using periodic uniform matrix product states},\
  }\href {https://doi.org/10.1103/PhysRevLett.121.230402} {\bibfield  {journal}
  {\bibinfo  {journal} {Phys. Rev. Lett.}\ }\textbf {\bibinfo {volume} {121}},\
  \bibinfo {pages} {230402} (\bibinfo {year} {2018})}\BibitemShut {NoStop}%
\bibitem [{\citenamefont {Pippan}\ \emph {et~al.}(2010)\citenamefont {Pippan},
  \citenamefont {White},\ and\ \citenamefont {Evertz}}]{Pippan2010}%
  \BibitemOpen
  \bibfield  {author} {\bibinfo {author} {\bibfnamefont {P.}~\bibnamefont
  {Pippan}}, \bibinfo {author} {\bibfnamefont {S.~R.}\ \bibnamefont {White}},\
  and\ \bibinfo {author} {\bibfnamefont {H.~G.}\ \bibnamefont {Evertz}},\
  }\bibfield  {title} {\bibinfo {title} {Efficient matrix-product state method
  for periodic boundary conditions},\ }\href
  {https://doi.org/10.1103/PhysRevB.81.081103} {\bibfield  {journal} {\bibinfo
  {journal} {Phys. Rev. B}\ }\textbf {\bibinfo {volume} {81}},\ \bibinfo
  {pages} {081103} (\bibinfo {year} {2010})}\BibitemShut {NoStop}%
\bibitem [{\citenamefont {Rossini}\ \emph {et~al.}(2011)\citenamefont
  {Rossini}, \citenamefont {Giovannetti},\ and\ \citenamefont
  {Fazio}}]{Rossini2011}%
  \BibitemOpen
  \bibfield  {author} {\bibinfo {author} {\bibfnamefont {D.}~\bibnamefont
  {Rossini}}, \bibinfo {author} {\bibfnamefont {V.}~\bibnamefont
  {Giovannetti}},\ and\ \bibinfo {author} {\bibfnamefont {R.}~\bibnamefont
  {Fazio}},\ }\bibfield  {title} {\bibinfo {title} {Stiffness in 1d matrix
  product states with periodic boundary conditions},\ }\href
  {https://doi.org/10.1088/1742-5468/2011/05/p05021} {\bibfield  {journal}
  {\bibinfo  {journal} {Journal of Statistical Mechanics: Theory and
  Experiment}\ }\textbf {\bibinfo {volume} {2011}},\ \bibinfo {pages} {P05021}
  (\bibinfo {year} {2011})}\BibitemShut {NoStop}%
\bibitem [{\citenamefont {Pirvu}\ \emph
  {et~al.}(2012{\natexlab{b}})\citenamefont {Pirvu}, \citenamefont {Haegeman},\
  and\ \citenamefont {Verstraete}}]{Pirvu2012a}%
  \BibitemOpen
  \bibfield  {author} {\bibinfo {author} {\bibfnamefont {B.}~\bibnamefont
  {Pirvu}}, \bibinfo {author} {\bibfnamefont {J.}~\bibnamefont {Haegeman}},\
  and\ \bibinfo {author} {\bibfnamefont {F.}~\bibnamefont {Verstraete}},\
  }\bibfield  {title} {\bibinfo {title} {Matrix product state based algorithm
  for determining dispersion relations of quantum spin chains with periodic
  boundary conditions},\ }\href {https://doi.org/10.1103/PhysRevB.85.035130}
  {\bibfield  {journal} {\bibinfo  {journal} {Phys. Rev. B}\ }\textbf {\bibinfo
  {volume} {85}},\ \bibinfo {pages} {035130} (\bibinfo {year}
  {2012}{\natexlab{b}})}\BibitemShut {NoStop}%
\bibitem [{\citenamefont {Tu}\ \emph {et~al.}(2021)\citenamefont {Tu},
  \citenamefont {Wu}, \citenamefont {Schuch}, \citenamefont {Kawashima},\ and\
  \citenamefont {Chen}}]{Chen2021}%
  \BibitemOpen
  \bibfield  {author} {\bibinfo {author} {\bibfnamefont {W.-L.}\ \bibnamefont
  {Tu}}, \bibinfo {author} {\bibfnamefont {H.-K.}\ \bibnamefont {Wu}}, \bibinfo
  {author} {\bibfnamefont {N.}~\bibnamefont {Schuch}}, \bibinfo {author}
  {\bibfnamefont {N.}~\bibnamefont {Kawashima}},\ and\ \bibinfo {author}
  {\bibfnamefont {J.-Y.}\ \bibnamefont {Chen}},\ }\bibfield  {title} {\bibinfo
  {title} {Generating function for tensor network diagrammatic summation},\
  }\href {https://arxiv.org/abs/2101.03935} {\bibfield  {journal} {\bibinfo
  {journal} {arXiv:2101.03935}\ } (\bibinfo {year} {2021})}\BibitemShut
  {NoStop}%
\bibitem [{\citenamefont {Stoudenmire}\ and\ \citenamefont
  {White}(2012)}]{Stoudenmire2012}%
  \BibitemOpen
  \bibfield  {author} {\bibinfo {author} {\bibfnamefont {E.~M.}\ \bibnamefont
  {Stoudenmire}}\ and\ \bibinfo {author} {\bibfnamefont {S.~R.}\ \bibnamefont
  {White}},\ }\bibfield  {title} {\bibinfo {title} {Studying two-dimensional
  systems with the density matrix renormalization group},\ }\href
  {https://doi.org/10.1146/annurev-conmatphys-020911-125018} {\bibfield
  {journal} {\bibinfo  {journal} {Annual Review of Condensed Matter Physics}\
  }\textbf {\bibinfo {volume} {3}},\ \bibinfo {pages} {111} (\bibinfo {year}
  {2012})}\BibitemShut {NoStop}%
\bibitem [{\citenamefont {Cincio}\ and\ \citenamefont
  {Vidal}(2013)}]{Cincio2013}%
  \BibitemOpen
  \bibfield  {author} {\bibinfo {author} {\bibfnamefont {L.}~\bibnamefont
  {Cincio}}\ and\ \bibinfo {author} {\bibfnamefont {G.}~\bibnamefont {Vidal}},\
  }\bibfield  {title} {\bibinfo {title} {Characterizing topological order by
  studying the ground states on an infinite cylinder},\ }\href
  {https://doi.org/10.1103/PhysRevLett.110.067208} {\bibfield  {journal}
  {\bibinfo  {journal} {Phys. Rev. Lett.}\ }\textbf {\bibinfo {volume} {110}},\
  \bibinfo {pages} {067208} (\bibinfo {year} {2013})}\BibitemShut {NoStop}%
\bibitem [{\citenamefont {Motruk}\ \emph {et~al.}(2016)\citenamefont {Motruk},
  \citenamefont {Zaletel}, \citenamefont {Mong},\ and\ \citenamefont
  {Pollmann}}]{Motruk2016}%
  \BibitemOpen
  \bibfield  {author} {\bibinfo {author} {\bibfnamefont {J.}~\bibnamefont
  {Motruk}}, \bibinfo {author} {\bibfnamefont {M.~P.}\ \bibnamefont {Zaletel}},
  \bibinfo {author} {\bibfnamefont {R.~S.~K.}\ \bibnamefont {Mong}},\ and\
  \bibinfo {author} {\bibfnamefont {F.}~\bibnamefont {Pollmann}},\ }\bibfield
  {title} {\bibinfo {title} {Density matrix renormalization group on a cylinder
  in mixed real and momentum space},\ }\href
  {https://doi.org/10.1103/PhysRevB.93.155139} {\bibfield  {journal} {\bibinfo
  {journal} {Phys. Rev. B}\ }\textbf {\bibinfo {volume} {93}},\ \bibinfo
  {pages} {155139} (\bibinfo {year} {2016})}\BibitemShut {NoStop}%
\bibitem [{\citenamefont {Ehlers}\ \emph {et~al.}(2017)\citenamefont {Ehlers},
  \citenamefont {White},\ and\ \citenamefont {Noack}}]{Ehlers2017}%
  \BibitemOpen
  \bibfield  {author} {\bibinfo {author} {\bibfnamefont {G.}~\bibnamefont
  {Ehlers}}, \bibinfo {author} {\bibfnamefont {S.~R.}\ \bibnamefont {White}},\
  and\ \bibinfo {author} {\bibfnamefont {R.~M.}\ \bibnamefont {Noack}},\
  }\bibfield  {title} {\bibinfo {title} {Hybrid-space density matrix
  renormalization group study of the doped two-dimensional hubbard model},\
  }\href {https://doi.org/10.1103/PhysRevB.95.125125} {\bibfield  {journal}
  {\bibinfo  {journal} {Phys. Rev. B}\ }\textbf {\bibinfo {volume} {95}},\
  \bibinfo {pages} {125125} (\bibinfo {year} {2017})}\BibitemShut {NoStop}%
\bibitem [{\citenamefont {{Van Damme}}\ \emph {et~al.}(2020)\citenamefont {{Van
  Damme}}, \citenamefont {Roose}, \citenamefont {Hauru},\ and\ \citenamefont
  {Haegeman}}]{mpskit}%
  \BibitemOpen
  \bibfield  {author} {\bibinfo {author} {\bibfnamefont {M.}~\bibnamefont {{Van
  Damme}}}, \bibinfo {author} {\bibfnamefont {G.}~\bibnamefont {Roose}},
  \bibinfo {author} {\bibfnamefont {M.}~\bibnamefont {Hauru}},\ and\ \bibinfo
  {author} {\bibfnamefont {J.}~\bibnamefont {Haegeman}},\ }\href
  {https://github.com/maartenvd/MPSKit.jl/} {\bibinfo {title} {Mpskit.jl}}
  (\bibinfo {year} {2020})\BibitemShut {NoStop}%
\bibitem [{\citenamefont {Schollw\"ock}(2005)}]{Schollwock2005}%
  \BibitemOpen
  \bibfield  {author} {\bibinfo {author} {\bibfnamefont {U.}~\bibnamefont
  {Schollw\"ock}},\ }\bibfield  {title} {\bibinfo {title} {The density-matrix
  renormalization group},\ }\href {https://doi.org/10.1103/RevModPhys.77.259}
  {\bibfield  {journal} {\bibinfo  {journal} {Rev. Mod. Phys.}\ }\textbf
  {\bibinfo {volume} {77}},\ \bibinfo {pages} {259} (\bibinfo {year}
  {2005})}\BibitemShut {NoStop}%
\bibitem [{\citenamefont {McCulloch}(2007)}]{McCulloch2007}%
  \BibitemOpen
  \bibfield  {author} {\bibinfo {author} {\bibfnamefont {I.~P.}\ \bibnamefont
  {McCulloch}},\ }\bibfield  {title} {\bibinfo {title} {{From density-matrix
  renormalization group to matrix product states}},\ }\href
  {https://doi.org/10.1088/1742-5468/2007/10/P10014} {\bibfield  {journal}
  {\bibinfo  {journal} {Journal of Statistical Mechanics: Theory and
  Experiment}\ ,\ \bibinfo {pages} {10014}} (\bibinfo {year}
  {2007})}\BibitemShut {NoStop}%
\bibitem [{\citenamefont {Chepiga}\ and\ \citenamefont
  {Mila}(2017)}]{Chepiga2017}%
  \BibitemOpen
  \bibfield  {author} {\bibinfo {author} {\bibfnamefont {N.}~\bibnamefont
  {Chepiga}}\ and\ \bibinfo {author} {\bibfnamefont {F.}~\bibnamefont {Mila}},\
  }\bibfield  {title} {\bibinfo {title} {Excitation spectrum and density matrix
  renormalization group iterations},\ }\href
  {https://doi.org/10.1103/PhysRevB.96.054425} {\bibfield  {journal} {\bibinfo
  {journal} {Phys. Rev. B}\ }\textbf {\bibinfo {volume} {96}},\ \bibinfo
  {pages} {054425} (\bibinfo {year} {2017})}\BibitemShut {NoStop}%
\bibitem [{\citenamefont {Vanderstraeten}\ \emph
  {et~al.}(2015{\natexlab{a}})\citenamefont {Vanderstraeten}, \citenamefont
  {Verstraete},\ and\ \citenamefont {Haegeman}}]{Vanderstraeten2015}%
  \BibitemOpen
  \bibfield  {author} {\bibinfo {author} {\bibfnamefont {L.}~\bibnamefont
  {Vanderstraeten}}, \bibinfo {author} {\bibfnamefont {F.}~\bibnamefont
  {Verstraete}},\ and\ \bibinfo {author} {\bibfnamefont {J.}~\bibnamefont
  {Haegeman}},\ }\bibfield  {title} {\bibinfo {title} {Scattering particles in
  quantum spin chains},\ }\href {https://doi.org/10.1103/PhysRevB.92.125136}
  {\bibfield  {journal} {\bibinfo  {journal} {Phys. Rev. B}\ }\textbf {\bibinfo
  {volume} {92}},\ \bibinfo {pages} {125136} (\bibinfo {year}
  {2015}{\natexlab{a}})}\BibitemShut {NoStop}%
\bibitem [{\citenamefont {White}\ and\ \citenamefont
  {Huse}(1993)}]{White1993b}%
  \BibitemOpen
  \bibfield  {author} {\bibinfo {author} {\bibfnamefont {S.~R.}\ \bibnamefont
  {White}}\ and\ \bibinfo {author} {\bibfnamefont {D.~A.}\ \bibnamefont
  {Huse}},\ }\bibfield  {title} {\bibinfo {title} {Numerical
  renormalization-group study of low-lying eigenstates of the antiferromagnetic
  s=1 heisenberg chain},\ }\href {https://doi.org/10.1103/PhysRevB.48.3844}
  {\bibfield  {journal} {\bibinfo  {journal} {Phys. Rev. B}\ }\textbf {\bibinfo
  {volume} {48}},\ \bibinfo {pages} {3844} (\bibinfo {year}
  {1993})}\BibitemShut {NoStop}%
\bibitem [{\citenamefont {S\o{}rensen}\ and\ \citenamefont
  {Affleck}(1993)}]{Sorensen1993}%
  \BibitemOpen
  \bibfield  {author} {\bibinfo {author} {\bibfnamefont {E.~S.}\ \bibnamefont
  {S\o{}rensen}}\ and\ \bibinfo {author} {\bibfnamefont {I.}~\bibnamefont
  {Affleck}},\ }\bibfield  {title} {\bibinfo {title} {Large-scale numerical
  evidence for bose condensation in the s=1 antiferromagnetic chain in a strong
  field},\ }\href {https://doi.org/10.1103/PhysRevLett.71.1633} {\bibfield
  {journal} {\bibinfo  {journal} {Phys. Rev. Lett.}\ }\textbf {\bibinfo
  {volume} {71}},\ \bibinfo {pages} {1633} (\bibinfo {year}
  {1993})}\BibitemShut {NoStop}%
\bibitem [{\citenamefont {Haegeman}\ and\ \citenamefont
  {Verstraete}(2017)}]{Haegeman2017}%
  \BibitemOpen
  \bibfield  {author} {\bibinfo {author} {\bibfnamefont {J.}~\bibnamefont
  {Haegeman}}\ and\ \bibinfo {author} {\bibfnamefont {F.}~\bibnamefont
  {Verstraete}},\ }\bibfield  {title} {\bibinfo {title} {Diagonalizing transfer
  matrices and matrix product operators: A medley of exact and computational
  methods},\ }\href {https://doi.org/10.1146/annurev-conmatphys-031016-025507}
  {\bibfield  {journal} {\bibinfo  {journal} {Annual Review of Condensed Matter
  Physics}\ }\textbf {\bibinfo {volume} {8}},\ \bibinfo {pages} {355} (\bibinfo
  {year} {2017})}\BibitemShut {NoStop}%
\bibitem [{\citenamefont {Zauner-Stauber}\ \emph
  {et~al.}(2018{\natexlab{b}})\citenamefont {Zauner-Stauber}, \citenamefont
  {Vanderstraeten}, \citenamefont {Fishman}, \citenamefont {Verstraete},\ and\
  \citenamefont {Haegeman}}]{ZaunerStauber2018}%
  \BibitemOpen
  \bibfield  {author} {\bibinfo {author} {\bibfnamefont {V.}~\bibnamefont
  {Zauner-Stauber}}, \bibinfo {author} {\bibfnamefont {L.}~\bibnamefont
  {Vanderstraeten}}, \bibinfo {author} {\bibfnamefont {M.~T.}\ \bibnamefont
  {Fishman}}, \bibinfo {author} {\bibfnamefont {F.}~\bibnamefont
  {Verstraete}},\ and\ \bibinfo {author} {\bibfnamefont {J.}~\bibnamefont
  {Haegeman}},\ }\bibfield  {title} {\bibinfo {title} {Variational optimization
  algorithms for uniform matrix product states},\ }\href
  {https://doi.org/10.1103/PhysRevB.97.045145} {\bibfield  {journal} {\bibinfo
  {journal} {Phys. Rev. B}\ }\textbf {\bibinfo {volume} {97}},\ \bibinfo
  {pages} {045145} (\bibinfo {year} {2018}{\natexlab{b}})}\BibitemShut
  {NoStop}%
\bibitem [{\citenamefont {Wang}\ and\ \citenamefont {Lin}(2019)}]{Wang2019}%
  \BibitemOpen
  \bibfield  {author} {\bibinfo {author} {\bibfnamefont {L.}~\bibnamefont
  {Wang}}\ and\ \bibinfo {author} {\bibfnamefont {H.-Q.}\ \bibnamefont {Lin}},\
  }\bibfield  {title} {\bibinfo {title} {Dynamic structure factor from real
  time evolution and exact correction vectors with matrix product states},\
  }\href {https://arxiv.org/abs/1901.07751} {\bibfield  {journal} {\bibinfo
  {journal} {arXiv:1901.07751}\ } (\bibinfo {year} {2019})}\BibitemShut
  {NoStop}%
\bibitem [{\citenamefont {Bl\"ote}\ and\ \citenamefont
  {Deng}(2002)}]{Blote2002}%
  \BibitemOpen
  \bibfield  {author} {\bibinfo {author} {\bibfnamefont {H.~W.~J.}\
  \bibnamefont {Bl\"ote}}\ and\ \bibinfo {author} {\bibfnamefont
  {Y.}~\bibnamefont {Deng}},\ }\bibfield  {title} {\bibinfo {title} {Cluster
  monte carlo simulation of the transverse ising model},\ }\href
  {https://doi.org/10.1103/PhysRevE.66.066110} {\bibfield  {journal} {\bibinfo
  {journal} {Phys. Rev. E}\ }\textbf {\bibinfo {volume} {66}},\ \bibinfo
  {pages} {066110} (\bibinfo {year} {2002})}\BibitemShut {NoStop}%
\bibitem [{\citenamefont {Qin}\ \emph {et~al.}(2021)\citenamefont {Qin},
  \citenamefont {Sch\"{a}fer}, \citenamefont {Andergassen}, \citenamefont
  {Corboz},\ and\ \citenamefont {Gull}}]{Qin2021}%
  \BibitemOpen
  \bibfield  {author} {\bibinfo {author} {\bibfnamefont {M.}~\bibnamefont
  {Qin}}, \bibinfo {author} {\bibfnamefont {T.}~\bibnamefont {Sch\"{a}fer}},
  \bibinfo {author} {\bibfnamefont {S.}~\bibnamefont {Andergassen}}, \bibinfo
  {author} {\bibfnamefont {P.}~\bibnamefont {Corboz}},\ and\ \bibinfo {author}
  {\bibfnamefont {E.}~\bibnamefont {Gull}},\ }\bibfield  {title} {\bibinfo
  {title} {The hubbard model: A computational perspective},\ }\href
  {https://arxiv.org/abs/2104.00064} {\bibfield  {journal} {\bibinfo  {journal}
  {arXiv:2104.00064}\ } (\bibinfo {year} {2021})}\BibitemShut {NoStop}%
\bibitem [{\citenamefont {Bultinck}\ \emph {et~al.}(2017)\citenamefont
  {Bultinck}, \citenamefont {Williamson}, \citenamefont {Haegeman},\ and\
  \citenamefont {Verstraete}}]{Bultinck2017}%
  \BibitemOpen
  \bibfield  {author} {\bibinfo {author} {\bibfnamefont {N.}~\bibnamefont
  {Bultinck}}, \bibinfo {author} {\bibfnamefont {D.~J.}\ \bibnamefont
  {Williamson}}, \bibinfo {author} {\bibfnamefont {J.}~\bibnamefont
  {Haegeman}},\ and\ \bibinfo {author} {\bibfnamefont {F.}~\bibnamefont
  {Verstraete}},\ }\bibfield  {title} {\bibinfo {title} {Fermionic matrix
  product states and one-dimensional topological phases},\ }\href
  {https://doi.org/10.1103/PhysRevB.95.075108} {\bibfield  {journal} {\bibinfo
  {journal} {Phys. Rev. B}\ }\textbf {\bibinfo {volume} {95}},\ \bibinfo
  {pages} {075108} (\bibinfo {year} {2017})}\BibitemShut {NoStop}%
\bibitem [{\citenamefont {Chiu}\ \emph {et~al.}(2019)\citenamefont {Chiu},
  \citenamefont {Ji}, \citenamefont {Bohrdt}, \citenamefont {Xu}, \citenamefont
  {Knap}, \citenamefont {Demler}, \citenamefont {Grusdt}, \citenamefont
  {Greiner},\ and\ \citenamefont {Greif}}]{Chiu2019}%
  \BibitemOpen
  \bibfield  {author} {\bibinfo {author} {\bibfnamefont {C.~S.}\ \bibnamefont
  {Chiu}}, \bibinfo {author} {\bibfnamefont {G.}~\bibnamefont {Ji}}, \bibinfo
  {author} {\bibfnamefont {A.}~\bibnamefont {Bohrdt}}, \bibinfo {author}
  {\bibfnamefont {M.}~\bibnamefont {Xu}}, \bibinfo {author} {\bibfnamefont
  {M.}~\bibnamefont {Knap}}, \bibinfo {author} {\bibfnamefont {E.}~\bibnamefont
  {Demler}}, \bibinfo {author} {\bibfnamefont {F.}~\bibnamefont {Grusdt}},
  \bibinfo {author} {\bibfnamefont {M.}~\bibnamefont {Greiner}},\ and\ \bibinfo
  {author} {\bibfnamefont {D.}~\bibnamefont {Greif}},\ }\bibfield  {title}
  {\bibinfo {title} {String patterns in the doped hubbard model},\ }\href
  {https://doi.org/10.1126/science.aav3587} {\bibfield  {journal} {\bibinfo
  {journal} {Science}\ }\textbf {\bibinfo {volume} {365}},\ \bibinfo {pages}
  {251} (\bibinfo {year} {2019})}\BibitemShut {NoStop}%
\bibitem [{\citenamefont {Bohrdt}\ \emph {et~al.}(2019)\citenamefont {Bohrdt},
  \citenamefont {Chiu}, \citenamefont {Ji}, \citenamefont {Xu}, \citenamefont
  {Greif}, \citenamefont {Greiner}, \citenamefont {Demler}, \citenamefont
  {Grusdt},\ and\ \citenamefont {Knap}}]{Bohrdt2019}%
  \BibitemOpen
  \bibfield  {author} {\bibinfo {author} {\bibfnamefont {A.}~\bibnamefont
  {Bohrdt}}, \bibinfo {author} {\bibfnamefont {C.~S.}\ \bibnamefont {Chiu}},
  \bibinfo {author} {\bibfnamefont {G.}~\bibnamefont {Ji}}, \bibinfo {author}
  {\bibfnamefont {M.}~\bibnamefont {Xu}}, \bibinfo {author} {\bibfnamefont
  {D.}~\bibnamefont {Greif}}, \bibinfo {author} {\bibfnamefont
  {M.}~\bibnamefont {Greiner}}, \bibinfo {author} {\bibfnamefont
  {E.}~\bibnamefont {Demler}}, \bibinfo {author} {\bibfnamefont
  {F.}~\bibnamefont {Grusdt}},\ and\ \bibinfo {author} {\bibfnamefont
  {M.}~\bibnamefont {Knap}},\ }\bibfield  {title} {\bibinfo {title}
  {{Classifying snapshots of the doped Hubbard model with machine learning}},\
  }\href {https://doi.org/10.1038/s41567-019-0565-x} {\bibfield  {journal}
  {\bibinfo  {journal} {Nature Physics}\ }\textbf {\bibinfo {volume} {15}},\
  \bibinfo {pages} {921} (\bibinfo {year} {2019})}\BibitemShut {NoStop}%
\bibitem [{\citenamefont {Béran}\ \emph {et~al.}(1996)\citenamefont {Béran},
  \citenamefont {Poilblanc},\ and\ \citenamefont {Laughlin}}]{Beran1997}%
  \BibitemOpen
  \bibfield  {author} {\bibinfo {author} {\bibfnamefont {P.}~\bibnamefont
  {Béran}}, \bibinfo {author} {\bibfnamefont {D.}~\bibnamefont {Poilblanc}},\
  and\ \bibinfo {author} {\bibfnamefont {R.}~\bibnamefont {Laughlin}},\
  }\bibfield  {title} {\bibinfo {title} {Evidence for composite nature of
  quasiparticles in the 2d t-j model},\ }\href
  {https://doi.org/https://doi.org/10.1016/0550-3213(96)00196-4} {\bibfield
  {journal} {\bibinfo  {journal} {Nuclear Physics B}\ }\textbf {\bibinfo
  {volume} {473}},\ \bibinfo {pages} {707} (\bibinfo {year}
  {1996})}\BibitemShut {NoStop}%
\bibitem [{\citenamefont {Laughlin}(1997)}]{Laughlin1997}%
  \BibitemOpen
  \bibfield  {author} {\bibinfo {author} {\bibfnamefont {R.~B.}\ \bibnamefont
  {Laughlin}},\ }\bibfield  {title} {\bibinfo {title} {Evidence for
  quasiparticle decay in photoemission from underdoped cuprates},\ }\href
  {https://doi.org/10.1103/PhysRevLett.79.1726} {\bibfield  {journal} {\bibinfo
   {journal} {Phys. Rev. Lett.}\ }\textbf {\bibinfo {volume} {79}},\ \bibinfo
  {pages} {1726} (\bibinfo {year} {1997})}\BibitemShut {NoStop}%
\bibitem [{\citenamefont {Bohrdt}\ \emph {et~al.}(2020)\citenamefont {Bohrdt},
  \citenamefont {Demler}, \citenamefont {Pollmann}, \citenamefont {Knap},\ and\
  \citenamefont {Grusdt}}]{Bohrdt2020}%
  \BibitemOpen
  \bibfield  {author} {\bibinfo {author} {\bibfnamefont {A.}~\bibnamefont
  {Bohrdt}}, \bibinfo {author} {\bibfnamefont {E.}~\bibnamefont {Demler}},
  \bibinfo {author} {\bibfnamefont {F.}~\bibnamefont {Pollmann}}, \bibinfo
  {author} {\bibfnamefont {M.}~\bibnamefont {Knap}},\ and\ \bibinfo {author}
  {\bibfnamefont {F.}~\bibnamefont {Grusdt}},\ }\bibfield  {title} {\bibinfo
  {title} {Parton theory of angle-resolved photoemission spectroscopy spectra
  in antiferromagnetic mott insulators},\ }\href
  {https://doi.org/10.1103/PhysRevB.102.035139} {\bibfield  {journal} {\bibinfo
   {journal} {Phys. Rev. B}\ }\textbf {\bibinfo {volume} {102}},\ \bibinfo
  {pages} {035139} (\bibinfo {year} {2020})}\BibitemShut {NoStop}%
\bibitem [{\citenamefont {Poilblanc}\ \emph {et~al.}(2016)\citenamefont
  {Poilblanc}, \citenamefont {Schuch},\ and\ \citenamefont
  {Affleck}}]{Poilblanc2016}%
  \BibitemOpen
  \bibfield  {author} {\bibinfo {author} {\bibfnamefont {D.}~\bibnamefont
  {Poilblanc}}, \bibinfo {author} {\bibfnamefont {N.}~\bibnamefont {Schuch}},\
  and\ \bibinfo {author} {\bibfnamefont {I.}~\bibnamefont {Affleck}},\
  }\bibfield  {title} {\bibinfo {title} {$\mathrm{SU}(2{)}_{1}$ chiral edge
  modes of a critical spin liquid},\ }\href
  {https://doi.org/10.1103/PhysRevB.93.174414} {\bibfield  {journal} {\bibinfo
  {journal} {Phys. Rev. B}\ }\textbf {\bibinfo {volume} {93}},\ \bibinfo
  {pages} {174414} (\bibinfo {year} {2016})}\BibitemShut {NoStop}%
\bibitem [{\citenamefont {Vanderstraeten}\ \emph {et~al.}(2020)\citenamefont
  {Vanderstraeten}, \citenamefont {Wybo}, \citenamefont {Chepiga},
  \citenamefont {Verstraete},\ and\ \citenamefont {Mila}}]{Vanderstraeten2020}%
  \BibitemOpen
  \bibfield  {author} {\bibinfo {author} {\bibfnamefont {L.}~\bibnamefont
  {Vanderstraeten}}, \bibinfo {author} {\bibfnamefont {E.}~\bibnamefont
  {Wybo}}, \bibinfo {author} {\bibfnamefont {N.}~\bibnamefont {Chepiga}},
  \bibinfo {author} {\bibfnamefont {F.}~\bibnamefont {Verstraete}},\ and\
  \bibinfo {author} {\bibfnamefont {F.}~\bibnamefont {Mila}},\ }\bibfield
  {title} {\bibinfo {title} {Spinon confinement and deconfinement in spin-1
  chains},\ }\href {https://doi.org/10.1103/PhysRevB.101.115138} {\bibfield
  {journal} {\bibinfo  {journal} {Phys. Rev. B}\ }\textbf {\bibinfo {volume}
  {101}},\ \bibinfo {pages} {115138} (\bibinfo {year} {2020})}\BibitemShut
  {NoStop}%
\bibitem [{\citenamefont {Vanderstraeten}\ \emph
  {et~al.}(2015{\natexlab{b}})\citenamefont {Vanderstraeten}, \citenamefont
  {Mari\"en}, \citenamefont {Verstraete},\ and\ \citenamefont
  {Haegeman}}]{Vanderstraeten2015b}%
  \BibitemOpen
  \bibfield  {author} {\bibinfo {author} {\bibfnamefont {L.}~\bibnamefont
  {Vanderstraeten}}, \bibinfo {author} {\bibfnamefont {M.}~\bibnamefont
  {Mari\"en}}, \bibinfo {author} {\bibfnamefont {F.}~\bibnamefont
  {Verstraete}},\ and\ \bibinfo {author} {\bibfnamefont {J.}~\bibnamefont
  {Haegeman}},\ }\bibfield  {title} {\bibinfo {title} {Excitations and the
  tangent space of projected entangled-pair states},\ }\href
  {https://doi.org/10.1103/PhysRevB.92.201111} {\bibfield  {journal} {\bibinfo
  {journal} {Phys. Rev. B}\ }\textbf {\bibinfo {volume} {92}},\ \bibinfo
  {pages} {201111} (\bibinfo {year} {2015}{\natexlab{b}})}\BibitemShut
  {NoStop}%
\bibitem [{\citenamefont {Vanderstraeten}\ \emph
  {et~al.}(2019{\natexlab{b}})\citenamefont {Vanderstraeten}, \citenamefont
  {Haegeman},\ and\ \citenamefont {Verstraete}}]{Vanderstraeten2019b}%
  \BibitemOpen
  \bibfield  {author} {\bibinfo {author} {\bibfnamefont {L.}~\bibnamefont
  {Vanderstraeten}}, \bibinfo {author} {\bibfnamefont {J.}~\bibnamefont
  {Haegeman}},\ and\ \bibinfo {author} {\bibfnamefont {F.}~\bibnamefont
  {Verstraete}},\ }\bibfield  {title} {\bibinfo {title} {Simulating excitation
  spectra with projected entangled-pair states},\ }\href
  {https://doi.org/10.1103/PhysRevB.99.165121} {\bibfield  {journal} {\bibinfo
  {journal} {Phys. Rev. B}\ }\textbf {\bibinfo {volume} {99}},\ \bibinfo
  {pages} {165121} (\bibinfo {year} {2019}{\natexlab{b}})}\BibitemShut
  {NoStop}%
\bibitem [{\citenamefont {Ponsioen}\ and\ \citenamefont
  {Corboz}(2020)}]{Ponsioen2020}%
  \BibitemOpen
  \bibfield  {author} {\bibinfo {author} {\bibfnamefont {B.}~\bibnamefont
  {Ponsioen}}\ and\ \bibinfo {author} {\bibfnamefont {P.}~\bibnamefont
  {Corboz}},\ }\bibfield  {title} {\bibinfo {title} {Excitations with projected
  entangled pair states using the corner transfer matrix method},\ }\href
  {https://doi.org/10.1103/PhysRevB.101.195109} {\bibfield  {journal} {\bibinfo
   {journal} {Phys. Rev. B}\ }\textbf {\bibinfo {volume} {101}},\ \bibinfo
  {pages} {195109} (\bibinfo {year} {2020})}\BibitemShut {NoStop}%
\end{thebibliography}%
